\documentclass[a4paper,12pt]{article}
\usepackage{amsmath}
\usepackage{graphicx,subfigure}
\usepackage{latexsym}
\usepackage{amsfonts}
\usepackage{amsthm}
\usepackage{bm}
\usepackage{epstopdf}
\usepackage{titlesec}
\newtheorem{thm}{Theorem}[section]

\newtheorem{rem}[thm]{Remark}

\title{Nonlocal Coupled HI-MKdV Systems}

\author{
Asl{\i} Pekcan \thanks{Email:aslipekcan@hacettepe.edu.tr} \\
{\small Department of Mathematics, Faculty of Science} \\
{\small Hacettepe University, 06800 Ankara - Turkey}
}

\date{\nonumber}
\begin{document}
\maketitle
\date{\nonumber}

\numberwithin{equation}{section}

\begin{abstract}
We first study coupled Hirota-Iwao modified KdV (HI-mKdV) systems and give all possible local and nonlocal reductions of them.
We then present Hirota bilinear forms of these systems and give one-soliton solutions of them with the help of pfaffians.
By using the soliton solutions of the coupled HI-mKdV systems for $N=2$ and $N=3$ we find one-soliton solutions of the local and nonlocal reduced equations.
For $N=4$ we obtain one-soliton solutions of the local and nonlocal reduced systems of two equations.\\

Korteweg-de Vries equations, Hirota bilinear form, Pfaffians, Soliton solutions.\\

\end{abstract}

\maketitle

\newpage\makeatletter
\def\l@paragraph{\@dottedtocline{4}{5.3em}{2.1em}}
\makeatother

\tableofcontents{}

\newpage

\section{Introduction}

In last decade, it has been shown by many researchers that by using different local and nonlocal symmetry reductions one can obtain local and
the time reflection symmetric (T-symmetric), the space reflection symmetric (S-symmetric), and the space-time reflection symmetric (ST-symmetric) nonlocal
equations from the general AKNS formalism \cite{AKNS} and other integrable hierarchies. The main aim in these works is to find new integrable equations and to obtain new interesting wave solutions. Particular examples are nonlocal nonlinear Schr\"{o}dinger (NLS) equation \cite{AbMu1}-\cite{jianke}, nonlocal real or complex modified Korteweg-de Vries (mKdV) equation \cite{AbMu2}, \cite{AbMu3}, \cite{GurPek2}-\cite{ma}, nonlocal real or complex sine-Gordon equation \cite{AbMu2}, \cite{AbMu3}, \cite{aflm}, nonlocal Fordy-Kulish equations \cite{GurPek3}, \cite{GursesFK}, nonlocal $N$-wave systems \cite{gerd2}, nonlocal multidimensional versions
of NLS \cite{fok}-\cite{gerd3}, and so on. The relations between local and nonlocal reductions are given in \cite{Vincent}, \cite{Yang}.

In this work,
we consider the coupled Hirota-Iwao modified Korteweg-de Vries (HI-mKdV) system \cite{IHT}, \cite{IH}
\begin{equation}\label{generaleqn}
\displaystyle \mu v_{i,t}+3(\sum_{j,k=1}^N c_{jk}v_jv_k)v_{i,x}+v_{i,xxx}=0
\end{equation}
for $i=1, 2, \ldots, N$ where $c_{jk}=c_{kj}$ and $\mu$ is a constant. Here we will study a type of (\ref{generaleqn}) given by
\begin{equation}\label{coupledmKdVN}
\displaystyle \mu v_{i,t}+3\rho v_{i,x}+v_{i,xxx}=0, \quad i=1, 2,\ldots, N,
\end{equation}
where
\begin{equation}\label{rho}
\rho=\sum_{j, k=1, \ldots, N } c_{jk}v_jv_k,
\end{equation}
with $c_{jk}=c_{kj}$ and $c_{jj}=0$.

We present soliton solutions of coupled HI-mKdV systems (\ref{coupledmKdVN}) for any $N$ expressed by pfaffians. We
 examine all local and nonlocal reductions of these systems in general. The $N=2$ case is equivalent to the case that we had studied in \cite{GurPek1}. For consistency, we
 also mention about this case. For $N=3$ we give all possible reductions and obtain reduced equations and systems of two equations. If we require one-soliton solution obtained
 by Type 1 approach \cite{GurPek2}, because of the yielding constraints, the systems of two equations reduce to single known equations. Since we do not have interesting results for
 $N=3$, we present the $N=3$ case in Appendix. We analyze the $N=4$ case in detail. By Ablowitz-Musslimani type reductions we obtain local and nonlocal systems of two equations.
 Different than the $N=3$ case, if we apply Type 1 approach to these reduced systems we still have local and nonlocal systems of two equations. We also give examples of one-soliton
 solutions for this case with the graphs of the solutions included.

\section{Local Reductions}
We define new sets of indices. Let the Greek indices $\alpha, \beta, \gamma, \ldots$ run from 1 to $M$ and let the Latin alphabet $a,b,c,\ldots $ runs from  $M+1$ to $N$. Then $v_{i}=(v_{\alpha}, v_{a})$ and we can write $ \rho$ defined in (\ref{rho}) as

\begin{equation}
\rho=c_{\alpha \beta}\, v_{\alpha}\, v_{\beta}+ 2c_{\alpha a}\, v_{\alpha} v_{a}+c_{ab}\, v_{a}\,v_{b},
\end{equation}
where $c_{jk}=c_{kj}$, $c_{jj}=0$, and repeated indices are summed up over their range. There are two types of local reductions.

\vspace{0.5cm}
\noindent
{\bf (a)} Let $v_{a}=k_{a}+A_{a \alpha}\,v_{\alpha}$ then the system (\ref{coupledmKdVN}) consistently reduces to

\begin{equation}\label{eqn1}
\mu v_{\alpha,t}+3 \rho \,v_{\alpha,x}+v_{\alpha,xxx}=0,~~\alpha=1,2,\ldots, M,
\end{equation}
where $k_{a}$ and $A_{a \alpha}$ are arbitrary constants and
\begin{equation}
\rho=q_{\alpha \beta}\, v_{\alpha} \, v_{\beta}+q_{\alpha} \, v_{\alpha}+q,
\end{equation}
where
\begin{eqnarray}
q&=&c_{ab}\, k_{a} k_{b}, \\
q_{\alpha}&=& 2k_{a}\, c_{a \alpha}+2k_{a}\, c_{ab} A_{b \alpha},~  \alpha=1,2,\ldots, M,\\
q_{\alpha \beta}&=&c_{\alpha \beta}+c_{a \alpha}\, A_{a \beta}+c_{a \beta}\, A_{a \alpha}+c_{a b}\,A_{a \alpha}\,A_{b \beta},~~\alpha, \beta=1,2,\ldots, M.
\end{eqnarray}

\vspace{0.5cm}
\noindent
{\bf (b)} Let $v_{a}=A_{a \alpha}\,\bar{v}_{\alpha}$ where a bar over a letter denotes complex conjugation. Then (\ref{coupledmKdVN}) consistently reduce to
\begin{equation}\label{eqn2}
\mu v_{\alpha,t}+3 \rho \,v_{\alpha,x}+v_{\alpha,xxx}=0,~~\alpha=1,2,\ldots, M,
\end{equation}
where
\begin{equation}
\rho=c_{\alpha \beta} v_{\alpha} v_{\beta}+2c_{\alpha a}\, A_{a \beta} v_{\alpha} \bar{v}_{\beta}+c_{ab} A_{a \alpha} A_{b \beta} \bar{v}_{\alpha}\, \bar{v}_{\beta}
\end{equation}
with the following constraints
\begin{eqnarray}
\bar{c}_{\alpha \beta}&=&c_{ab}\, A_{a \alpha}\, A_{b \beta},~ \alpha, \beta=1, 2, \ldots, M,\\
c_{\alpha a}\, A_{a \beta}&=&\bar{c}_{\beta b} \bar{A}_{b \alpha},~ \alpha, \beta=1,2, \ldots, M,
\end{eqnarray}
so that $\rho=\bar{\rho}$.

\section{Nonlocal Reductions}

Let $v_{\alpha}^{\varepsilon}=v_{\alpha}(\varepsilon_{1} t, \varepsilon_{2} x)$ where $\varepsilon_{1}^2=\varepsilon_{2}^2=1$. Then there are also two different types of nonlocal reductions.

\vspace{0.5cm}
\noindent
{\bf (a)} Let $v_{a}=A_{a \alpha}\,v_{\alpha}^{\varepsilon}$, then (\ref{coupledmKdVN}) consistently reduce to

\begin{equation}\label{eqn3}
\mu v_{\alpha,t}+3 \rho \,v_{\alpha,x}+v_{\alpha,xxx}=0,~~\alpha=1, 2, \ldots, M,
\end{equation}
where
\begin{equation}
\rho=c_{\alpha \beta} v_{\alpha} v_{\beta}+ 2c_{\alpha a} A_{a \beta} v_{\alpha} v_{\beta}^{\varepsilon}+c_{ab}\,A_{a \beta} A_{b \alpha} v_{\alpha}^{\varepsilon}\,v_{\beta}^{\varepsilon}
\end{equation}
with the constraints
\begin{eqnarray}
c_{\alpha \beta}&=&c_{ab} A_{b \alpha}\,A_{a \beta},~  \alpha, \beta=1, 2, \ldots, M,\\
c_{\alpha a}A_{a \beta}&=&c_{\beta b}\,A_{b \alpha}, ~ \alpha, \beta=1, 2, \ldots, M,\\
\varepsilon_{1} \varepsilon_{2}&=&1,
\end{eqnarray}
so that $\rho=\rho^{\varepsilon}$.

\vspace{0.5cm}
\noindent
{\bf (b)} Let $v_{a}=A_{a \alpha}\,\bar{v}_{\alpha}^{\varepsilon}$, then (\ref{coupledmKdVN}) consistently reduce to
\begin{equation}\label{eqn4}
\mu v_{\alpha,t}+3 \rho \,v_{\alpha,x}+v_{\alpha,xxx}=0,~~\alpha=1, 2, \ldots, M,
\end{equation}
where
\begin{equation}
\rho=c_{\alpha \beta} v_{\alpha} v_{\beta}+ 2c_{\alpha a} A_{a \beta} v_{\alpha} \bar{v}_{\beta}^{\varepsilon}+c_{ab}\,A_{a \alpha} a_{b \beta} \bar{v}_{\alpha}^{\varepsilon}\,\bar{v}_{\beta}^{\varepsilon}
\end{equation}
with the constraints
\begin{eqnarray}
 \bar{c}_{\alpha \beta}&=&c_{ab} A_{b \alpha}\,A_{a \beta}, ~ \alpha, \beta=1, 2, \ldots, M,\\
c_{\alpha a}A_{a \beta}&=&\bar{c}_{\beta b}\bar{A}_{b \alpha}, ~\alpha, \beta=1, 2 ,\ldots, M,\\
\bar{\mu} \varepsilon_{1} \varepsilon_{2}&=&\mu,
\end{eqnarray}
so that $\rho=\bar{\rho}^{\varepsilon}$.

\section{Hirota Bilinear Method for Coupled HI-mKdV\\ Systems}

Let
\begin{equation}
\displaystyle v_i=\frac{g_i}{f}, \quad i=1, 2, \ldots, N
\end{equation}
in (\ref{coupledmKdVN}). Then Hirota bilinear form of (\ref{coupledmKdVN}) can be found as
\begin{align}
&(\mu D_t+D_x^2-3\lambda_i )\{g_i\cdot f\}=0,\\
&(D_x^2-\lambda_i)\{f\cdot f\}=\sum_{j, k=1, \ldots, N } c_{jk}g_jg_k.
\end{align}
For simplicity let us take $\lambda_i=0$, $i=1, 2, \ldots, N$. Hence we have
\begin{align}
&(\mu D_t+D_x^2)\{g_i\cdot f\}=0, \, i=1, 2,\ldots, N, \\
&D_x^2\{f\cdot f\}=\sum_{j, k=1, \ldots, N } c_{jk}g_jg_k
\end{align}
as Hirota bilinear form of (\ref{coupledmKdVN}). In \cite{IH}, multi-soliton solution of (\ref{coupledmKdVN})
which has $M_i$ solitons for $v_i$, $i=1,2,\ldots, N$, respectively is expressed by pfaffians as
\begin{align}\label{g_if}
&g_i=\mathrm{pf}(d_0,a_1,\ldots,a_L,b_1,\ldots,b_L,\beta_i)\\
&f=\mathrm{pf}(a_1,\ldots,a_L,b_1,\ldots,b_L)
\end{align}
for $i=1,2,\ldots, N$ and $L=M_1+M_2+\ldots+M_N$. Here the elements of pfaffians are defined as
\begin{align}
&\mathrm{pf}(d_n,a_i)=\frac{\partial^n}{\partial x^n}e^{\theta_i}=k_i^ne^{\theta_i},\, \theta_i=k_ix-k_i^3t+\alpha_i^0\nonumber\\
&\mathrm{pf}(a_i,a_m)=\frac{k_i-k_m}{k_i+k_m}e^{\theta_i+\theta_m}\nonumber\\
&\mathrm{pf}(b_i,b_m)=-\frac{c_{jk}}{k_i^2-k_m^2},\, b_i\in B_j,\, b_m\in B_k \label{pf1}
\end{align}
\begin{align}\label{pf2}
\mathrm{pf}(a_i,b_m)=\left\{ \begin{array}{cc}
                1, & \hspace{5mm} \mathrm{if}\quad  i=m \\
                0, & \hspace{5mm} \mathrm{if}\quad i\neq m
                                \end{array} \right.
\end{align}
\begin{align}\label{pf3}
\mathrm{pf}(b_m,\beta_i)=\left\{ \begin{array}{cc}
                1, & \hspace{5mm} \mathrm{if}\quad b_m\in B_i \\
                0, & \hspace{5mm} \mathrm{if}\quad b_m \notin B_i
\end{array} \right.
\end{align}
and $\mathrm{pf}(\mathrm{otherwise})=0$. Here the class of the sets $B_i$ of letters chosen out of $\{b_1,\ldots, b_L\}$, $i=1, 2, \ldots, N$, satisfies the following condition
\begin{align}
&M_i=\mathrm{number}\, \mathrm{of}\, \mathrm{elements}\, \mathrm{in}\, \mathrm{the}\, \mathrm{set} B_i\nonumber\\
& B_i \bigcap B_j=\emptyset,\,\mathrm{if}\, i\neq j\nonumber\\
&\displaystyle \bigcup\limits_{i=1}^N B_i=\{b_1,\ldots,b_L \}.\label{pf4}
\end{align}

\section{N=2 Coupled HI-mKdV System}

The system (\ref{coupledmKdVN}) for $N=2$ is
\begin{align}
& \mu v_{1,t}+3\rho v_{1,x}+v_{1,xxx}=0,\label{N=2a}\\
& \mu v_{2,t}+3\rho v_{2,x}+v_{2,xxx}=0,\label{N=2b}
\end{align}
where
\begin{equation}
\rho= 2c_{12}v_1v_2.
\end{equation}
The corresponding Hirota bilinear form is
\begin{align}
&(\mu D_t+D_x^3)\{g_i\cdot f\}=0,\quad i=1, 2, \\
&D_x^2\{f\cdot f\}= 2c_{12}g_1g_2.
\end{align}

\subsection{One-Soliton Solution of N=2 Coupled HI-mKdV System}

Here we will consider the solution given by (\ref{g_if}) with the pfaffian elements (\ref{pf1})-(\ref{pf3}) under the condition (\ref{pf4}).
The solution which has one-soliton for every $\displaystyle v_i=\frac{g_i}{f}$, $i=1, 2$, i.e., $M_1=1$, $M_2=1$ so $L=M_1+M_2=2$ with
$B_1=\{b_1\}$ and $B_2=\{b_2\}$ is expressed by
\begin{align}
&g_i=\mathrm{pf}(d_0,a_1,a_2,b_1,b_2,\beta_i),\, i=1,2,\\
&f=\mathrm{pf}(a_1,a_2,b_1,b_2),
\end{align}
which are explicitly given as
\begin{align}
& g_i=-e^{\theta_i},\quad \theta_i=k_ix-\frac{k_i^3}{a}t+\delta_i,\,\, i=1, 2,\\
&f=-1-\frac{c_{12}}{(k_1+k_2)^2}e^{\theta_1+\theta_2}.
\end{align}
Hence one-soliton solution of the system (\ref{N=2a})-(\ref{N=2b}) is
\begin{equation}
\displaystyle v_i=\frac{e^{\theta_i}}{1+\frac{c_{12}}{(k_1+k_2)^2}e^{\theta_1+\theta_2}},\,\, i=1,2.
\end{equation}
Here $k_i, \delta_i$, $i=1, 2$ are arbitrary constants. The above solution is exactly the solution we obtained in \cite{GurPek2} for $c_{12}=-1$.

\section{Local and Nonlocal Reductions for N=2}

The coupled HI-mKdV system for $N=2$ given by (\ref{N=2a})-(\ref{N=2b}) has four consistent reductions; two of them are local and the others are nonlocal.
To obtain one-soliton solution of the reduced equations we use Type 1 and Type 2 approaches given in \cite{GurPek2}. Since the solutions for $N=2$ case
are given in \cite{GurPek2} we will not present them here.

\subsection{Local Reductions for N=2}

The coupled HI-mKdV system for $N=2$ given by (\ref{N=2a})-(\ref{N=2b}) has two local reductions.
To obtain one-soliton solution of the reduced equations we use Type 1 and Type 2 approaches \cite{GurPek2}.\\

\noindent \textbf{i.} $v_2=a_2+a_1v_1$, $a_i$, $i=1, 2$ are constants. When we apply the reduction to
the equation (\ref{N=2b}) we obtain the equation (\ref{N=2a})
\begin{equation}
\mu v_{1,t}+3\rho v_{1,x}+v_{1,xxx}=0,
\end{equation}
without any additional condition on the parameters.
Hence the reduction is consistent and the reduced equation is
\begin{equation}\label{localiN=2}
\mu v_{1,t}+6\alpha v_1v_{1,x}+6\beta v_1^2v_{1,x}+v_{1,xxx}=0
\end{equation}
for $\alpha=c_{12}a_2$ and $\beta=c_{12}a_1$. This equation is a combination of KdV and mKdV equations.

\noindent If we follow the Type 1 approach, the constraints that the parameters of the one-soliton solution of (\ref{localiN=2}) are obtained from
\begin{equation}
\frac{g_2}{f}=a_2+a_1\frac{g_1}{f} \Rightarrow g_2=a_2f+a_1g_1,
\end{equation}
 as
\begin{equation}
1)\, a_2=0,\quad 2)\, k_2=k_1,\quad 3)\, e^{\delta_2}=a_1e^{\delta_1}.
\end{equation}

\vspace{0.7cm}

\noindent \textbf{ii.} $v_2=a_2+a_1\bar{v}_1$, $a_i$, $i=1, 2$ are constants. We use this reduction in (\ref{N=2b}) and obtain the following
conditions to have consistent reduction:
\begin{equation}\label{condlocaliiN=2}
1)\,\,  \mu=\bar{\mu},\,\, 2)\,\,  a_2=0, \,\, 3)\,\,  c_{12}a_1=\bar{c}_{12}\bar{a}_1.
\end{equation}
 So the reduced equation is the complex mKdV (cmKdV) equation
 \begin{equation}\label{localiiN=2}
\mu v_{1,t}+6\alpha |v_1|^2v_{1,x}+v_{1,xxx}=0,
\end{equation}
where $\alpha=c_{12}a_1$ and $\mu$ are real numbers.

\noindent From the Type 1 approach, the constraints that the parameters of the one-soliton solution of (\ref{localiiN=2}) are obtained from
\begin{equation}
\frac{g_2}{f}=a_1\frac{\bar{g}_1}{\bar{f}} \Rightarrow g_2=a_1\bar{g}_1,\quad f=\bar{f}
\end{equation}
 as
\begin{equation}
1)\, k_2=\bar{k}_1,\quad 2)\, e^{\delta_2}=a_1e^{\bar{\delta}_1},
\end{equation}
besides the conditions (\ref{condlocaliiN=2}).

\subsection{Nonlocal Reductions for N=2}

\noindent \textbf{i.} \textbf{$v_2(t,x)=a_2+a_1v_1(\varepsilon_1t,\varepsilon_2 x)=a_2+a_1v_1^{\varepsilon}$}, $\varepsilon_i^2=1$, $a_i$, $i=1, 2$ are constants. When we use this reduction
in (\ref{N=2b}) for consistency of reduction we get the following conditions:
\begin{equation}
1)\, \varepsilon_1\varepsilon_2=1,\quad 2)\, a_2=0.
\end{equation}
Therefore to have a nonlocal equation, there is only one possibility; $(\varepsilon_1,\varepsilon_2)=(-1,-1)$. The reduced equation is ST-symmetric nonlocal mKdV equation,
\begin{equation}\label{nonlocaliN=2}
\mu v_{1,t}(t,x)+6\alpha v_1(t,x)v_1(-t,-x)v_{1,x}(t,x)+v_{1,xxx}(t,x)=0,
\end{equation}
where $\alpha=c_{12}a_1$.

\noindent In this case, if we use the Type 1 approach to find one-soliton solution of (\ref{nonlocaliN=2}) we get trivial solution.
Hence we use the Type 2 approach. Therefore the constraints that the parameters of the one-soliton solution of (\ref{localiiN=2}) are obtained from
\begin{equation}
g_2f^{\varepsilon}-a_1fg_1^{\varepsilon}=0\,\, \mathrm{where}\,\,  g_1^{\varepsilon}=g_1(-t,-x),\,\, f^{\varepsilon}=f(-t,-x),
\end{equation}
 as
\begin{equation}
1)\, e^{\delta_1}=\pm\frac{(k_1+k_2)}{\sqrt{c_{12}a_1}} ,\quad 2)\, e^{\delta_2}=\pm\frac{\sqrt{a_1}(k_1+k_2)}{\sqrt{c_{12}}}.
\end{equation}

\bigskip

\noindent \textbf{ii.} \textbf{$v_2(t,x)=a_2+a_1\bar{v}_1(\varepsilon_1t,\varepsilon_2 x)=a_2+a_1\bar{v}_1^{\varepsilon}$}, $\varepsilon_i^2=1$, $a_i$, $i=1, 2$ are constants. Applying this reduction to (\ref{N=2b}) gives the constraints on the parameters as
\begin{equation}\label{condnonlocaliiN=2}
1)\, \mu=\bar{\mu}\varepsilon_1\varepsilon_2,\quad 2)\, a_2=0,\quad 3)\, c_{12}a_1=\bar{c}_{12}\bar{a}_1.
\end{equation}
Hence we have three different nonlocal cmKdV equations:\\

\noindent \textbf{a.} \textbf{T-Symmetric Nonlocal CMKdV Equation:}
\begin{equation}\label{TsymmN=2}
\mu v_{1,t}(t,x)+6\alpha v_1(t,x)\bar{v}_1(-t,x)+v_{1,xxx}(t,x)=0,
\end{equation}
where $\mu=-\bar{\mu}$ and $\alpha=c_{12}a_1\in \mathbb{R}$.\\

\noindent \textbf{b.} \textbf{S-Symmetric Nonlocal CMKdV Equation:}
\begin{equation}\label{SsymmN=2}
\mu v_{1,t}(t,x)+6\alpha v_1(t,x)\bar{v}_1(t,-x)+v_{1,xxx}(t,x)=0,
\end{equation}
where $\mu=-\bar{\mu}$ and $\alpha=c_{12}a_1\in \mathbb{R}$.\\

\noindent \textbf{c.} \textbf{ST-Symmetric Nonlocal CMKdV Equation:}
\begin{equation}\label{STsymmN=2}
\mu v_{1,t}(t,x)+6\alpha v_1(t,x)\bar{v}_1(-t,-x)+v_{1,xxx}(t,x)=0,
\end{equation}
where $\mu=\bar{\mu}$ and $\alpha=c_{12}a_1\in \mathbb{R}$.\\

\noindent If we use the Type 1 approach, we get the constraints that the parameters of the one-soliton solutions of (\ref{TsymmN=2})-(\ref{STsymmN=2}) from
\begin{equation}
g_2=a_1\bar{g}_1^{\varepsilon},\,\, f=\bar{f}^{\varepsilon}\,\, \mathrm{where}\,\, \bar{g}_1^{\varepsilon}=\bar{g}_1(\varepsilon_1 t,\varepsilon_2 x),\,
\bar{f}^{\varepsilon}=\bar{f}(\varepsilon_1 t,\varepsilon_2 x),\,\, \varepsilon_i^2=1,\,\, i=1, 2,
\end{equation}
as
\begin{equation}
1)\, k_2=\varepsilon_2 \bar{k}_1,\quad 2)\, e^{\delta_2}=a_1e^{\delta_1},
\end{equation}
besides the conditions (\ref{condnonlocaliiN=2}).

 \begin{rem}
 Note that since we have given one-soliton solutions of the local and nonlocal reduced equations for $N=2$ in \cite{GurPek2}, we are not presenting them here also.
\end{rem}

\section{N=4 Coupled HI-mKdV System}

The system (\ref{coupledmKdVN}) for $N=4$ is
\begin{align}
& \mu v_{1,t}+3\rho v_{1,x}+v_{1,xxx}=0,\label{N=4a}\\
& \mu v_{2,t}+3\rho v_{2,x}+v_{2,xxx}=0,\label{N=4b}\\
& \mu v_{3,t}+3\rho v_{3,x}+v_{3,xxx}=0,\label{N=4c}\\
& \mu v_{4,t}+3\rho v_{4,x}+v_{4,xxx}=0,\label{N=4d}
\end{align}
where
\begin{equation}
\rho= 2(c_{12}v_1v_2+c_{13}v_1v_3+c_{14}v_1v_4+c_{23}v_2v_3+c_{24}v_2v_4+c_{34}v_3v_4),
\end{equation}
and $\mu$ is a constant.
The corresponding Hirota bilinear form is
\begin{align}
&(\mu D_t+D_x^3)\{g_i\cdot f\}=0,\quad i=1, 2, 3, 4,\\
&D_x^2\{f\cdot f\}= 2(c_{12}g_1g_2+c_{13}g_1g_3+c_{14}g_1g_4+c_{23}g_2g_3+c_{24}g_2g_4+c_{34}g_3g_4).
\end{align}

\subsection{One-Soliton Solution of N=4 Coupled HI-mKdV System}

If we consider the solution given by (\ref{g_if}) with the pfaffian elements (\ref{pf1})-(\ref{pf3}) under the condition (\ref{pf4}),
the solution which has one-soliton for every $\displaystyle v_i=\frac{g_i}{f}$ i.e., $M_i=1$, $i=1, 2, 3, 4$ so $L=M_1+M_2+M_3+M_4=4$ with
$B_i=\{b_i\}$, $i=1, 2, 3, 4$ is expressed by
\begin{align}
&g_i=\mathrm{pf}(d_0,a_1,a_2,a_3,a_4,b_1,b_2,b_3,b_4,\beta_i),\, i=1, 2, 3, 4,\\
&f=\mathrm{pf}(a_1,a_2,a_3,a_4,b_1,b_2,b_3,b_4),
\end{align}
which are explicitly written as
\begin{align}
& g_1=e^{\theta_1}+\alpha_{12}\alpha_{13}\alpha_{23}\beta_{23}e^{\theta_1+\theta_2+\theta_3}
+\alpha_{12}\alpha_{14}\alpha_{24}\beta_{24}e^{\theta_1+\theta_2+\theta_4}+\alpha_{13}\alpha_{14}\alpha_{34}\beta_{34}e^{\theta_1+\theta_3+\theta_4},\\
& g_2=e^{\theta_2}-\alpha_{12}\alpha_{13}\alpha_{23}\beta_{13}e^{\theta_1+\theta_2+\theta_3}-
\alpha_{12}\alpha_{14}\alpha_{24}\beta_{14}e^{\theta_1+\theta_2+\theta_4}+\alpha_{23}\alpha_{24}\alpha_{34}\beta_{34}e^{\theta_2+\theta_3+\theta_4},\\
& g_3=e^{\theta_3}+\alpha_{12}\alpha_{13}\alpha_{23}\beta_{12}e^{\theta_1+\theta_2+\theta_3}
-\alpha_{13}\alpha_{14}\alpha_{34}\beta_{14}e^{\theta_1+\theta_3+\theta_4}-\alpha_{23}\alpha_{24}\alpha_{34}\beta_{24}e^{\theta_2+\theta_3+\theta_4},\\
&g_4=e^{\theta_4}+\alpha_{12}\alpha_{14}\alpha_{24}\beta_{12}e^{\theta_1+\theta_2+\theta_4}+\alpha_{13}\alpha_{14}\alpha_{34}\beta_{13}e^{\theta_1+\theta_3+\theta_4}
+\alpha_{23}\alpha_{24}\alpha_{34}\beta_{23}e^{\theta_2+\theta_3+\theta_4},\\
&f=1+\alpha_{12}\beta_{12}e^{\theta_1+\theta_2}+\alpha_{13}\beta_{13}e^{\theta_1+\theta_3}+\alpha_{14}\beta_{14}e^{\theta_1+\theta_4}
+\alpha_{23}\beta_{23}e^{\theta_2+\theta_3}+\alpha_{24}\beta_{24}e^{\theta_2+\theta_4}\nonumber\\
&\hspace{1cm}+\alpha_{34}\beta_{34}e^{\theta_3+\theta_4}+\alpha_{12}\alpha_{13}\alpha_{14}\alpha_{23}\alpha_{24}\alpha_{34}[\beta_{12}\beta_{34}-\beta_{13}\beta_{24}
+\beta_{14}\beta_{23}]e^{\theta_1+\theta_2+\theta_3+\theta_4},\nonumber\\
\end{align}
where $\theta_i=k_ix-\frac{k_i^3}{\mu}t+\delta_i$, $\displaystyle \alpha_{ij}=\frac{k_i-k_j}{k_i+k_j}$, and $\displaystyle \beta_{ij}=\frac{c_{ij}}{k_i^2-k_j^2}$ for
$i,j=1,2,3,4$. Here $k_i, \delta_i$, $i=1, 2, 3, 4$ are arbitrary constants.

\section{Local and Nonlocal Reductions for N=4}
The coupled HI-mKdV system for $N=4$ given by (\ref{N=4a})-(\ref{N=4d}) has four consistent reductions; two of them are local and the others are nonlocal.
To obtain one-soliton solution of the reduced equations we use Type 1 and Type 2 approaches given in \cite{GurPek2}.

\subsection{Local Reductions for N=4}

Here we present two different local reductions of the coupled HI-mKdV system (\ref{coupledmKdVN}) for $N=4$.

\noindent \textbf{i.} \textbf{$v_4=a_1v_2$, $v_3=b_1v_1$}, $a_1$ and $b_1$ are constants. The system (\ref{N=4a})-(\ref{N=4d}) is
reduced to a system of two equations without any additional condition by this reduction. The reduced system is
\begin{align}
&\mu v_{1,t}+6[(c_{12}+c_{14}a_1+c_{23}b_1+c_{34}a_1b_1)v_1v_2+c_{24}a_1v_2^2+c_{13}b_1v_1^2]v_{1,x}+v_{1,xxx}=0\\
&\mu v_{2,t}+6[(c_{12}+c_{14}a_1+c_{23}b_1+c_{34}a_1b_1)v_1v_2+c_{24}a_1v_2^2+c_{13}b_1v_1^2]v_{2,x}+v_{2,xxx}=0.\label{localiN=4}
\end{align}
If we use the Type 1 approach, the constraints for the parameters of one-soliton solution of the system (\ref{localiN=4}) are found from the following equalities:
\begin{align}
\displaystyle
&v_4=a_1v_2 \Rightarrow \frac{g_4}{f}=a_1\frac{g_2}{f} \Rightarrow g_4=a_1g_2,\\
&v_3=b_1v_1 \Rightarrow \frac{g_3}{f}=b_1\frac{g_1}{f} \Rightarrow g_3=b_1g_1.
\end{align}
To satisfy the above equalities one of the cases for the parameters is
\begin{equation}
1)\, k_4=k_2, \quad 2)\, k_3=k_1,\quad 3)\, e^{\delta_4}=a_1e^{\delta_2},\quad 4)\, e^{\delta_3}=b_1e^{\delta_1}.
\end{equation}
Under these constraints, one-soliton solution of the system (\ref{localiN=4}) becomes
\begin{equation}\displaystyle
v_1=\frac{g_1}{f},\quad v_2=\frac{g_2}{f}
\end{equation}
where
\begin{align}
&g_1=e^{k_1x-\frac{k_1^3}{\mu}t+\delta_1}+\frac{a_1c_{24}(k_1-k_2)^2}{4k_2^2(k_1+k_2)^2}e^{(k_1+2k_2)x-\frac{(k_1^3+2k_2^3)}{\mu}t+\delta_1+2\delta_2}\\
&g_2=e^{k_2x-\frac{k_2^3}{\mu}t+\delta_1}+\frac{b_1c_{13}(k_1-k_2)^2}{4k_1^2(k_1+k_2)^2}e^{(2k_1+k_2)x-\frac{(2k_1^3+k_2^3)}{\mu}t+2\delta_1+\delta_2}
\end{align}
and
\begin{align}
f=&1+\frac{1}{(k_1+k_2)^2}[c_{12}+a_1c_{14}+b_1c_{23}+a_1b_1c_{34}]e^{(k_1+k_2)x-\frac{(k_1^3+k_2^3)}{\mu}t+\delta_1+\delta_2}\nonumber\\
&+\frac{b_1c_{13}}{4k_1^2}e^{2k_1x-\frac{2k_1^3}{\mu}t+2\delta_1}
+\frac{a_1c_{24}}{4k_2^2}e^{2k_2x-\frac{2k_2^3}{\mu}t+2\delta_2}\nonumber\\
&+\frac{a_1b_1c_{13}c_{24}(k_1-k_2)^4}{16k_1^2k_2^2(k_1+k_2)^4}e^{(2k_1+2k_2)x-\frac{(2k_1^3+2k_2^3)}{\mu}t+2\delta_1+2\delta_2}.
\end{align}

\noindent Consider the following example.\\

\noindent \textbf{Example 1.} Choose the parameters as $\mu=3, k_1=\frac{1}{2}, k_2=\frac{3}{2}$, and $a_1=b_1=e^{\delta_1}=e^{\delta_2}=c_{ij}=1$, $1\leq i< j\leq 4$. Hence the pair of the solution of the system (\ref{localiN=4}) becomes
\begin{align}\displaystyle
&v_1=\frac{4e^{\frac{1}{2}x+\frac{85}{24}t}(36e^{\frac{9}{4}t}+e^{3x})}{144e^{\frac{35}{6}t}+16e^{3x+\frac{43}{12}t}+144e^{2x+\frac{14}{3}t}
+144e^{x+\frac{23}{4}t}+e^{4x+\frac{7}{2}t}}\\
&v_2=\frac{36e^{\frac{3}{2}x+\frac{37}{8}t}(4e^{\frac{1}{12}t}+e^{x})}{144e^{\frac{35}{6}t}+16e^{3x+\frac{43}{12}t}+144e^{2x+\frac{14}{3}t}
+144e^{x+\frac{23}{4}t}+e^{4x+\frac{7}{2}t}}.
\end{align}
These solutions are finite and bounded for $t\geq 0$. The graphs of the functions $v_1$ and $v_2$ are given in Figure 1(a) and Figure 1(b) respectively.
\begin{figure}[h!]
\centering     
\subfigure[]{\label{fig:locali1a}\includegraphics[width=26mm]{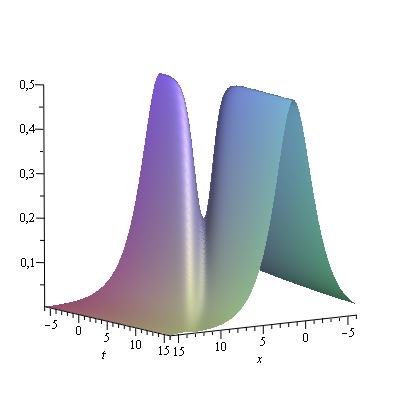}}\hspace{5cm}
\subfigure[]{\label{fig:locali1a}\includegraphics[width=26mm]{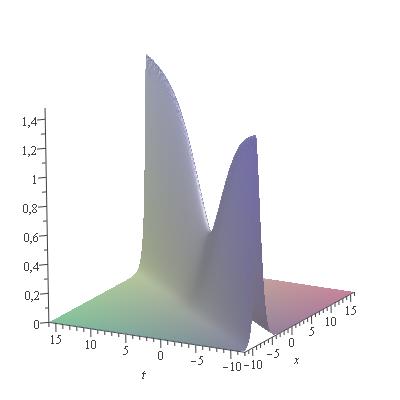}}
\caption{One-soliton solution of the system (\ref{localiN=4}) with the parameters $\mu=3, k_1=\frac{1}{2}, k_2=\frac{3}{2}, a_1=b_1=e^{\delta_1}=e^{\delta_2}=c_{ij}=1, 1\leq i< j\leq 4$.}
\end{figure}

\bigskip

\noindent \textbf{ii.} $v_4=a_1\bar{v}_2$, $v_3=b_1\bar{v}_1$, $a_1$ and $b_1$ are constants. To have a consistent reduction of the
system (\ref{N=4a})-(\ref{N=4d}) we must have
\begin{equation}
 \mu=\bar{\mu},\quad  \rho=\bar{\rho}.
\end{equation}
The relation $\rho=\bar{\rho}$ is satisfied when
\begin{equation}\label{N=4localiicond}
1)\, c_{12}=\bar{c}_{34}\bar{a}_1\bar{b}_1,\quad 2)\, c_{14}a_1=\bar{c}_{23}\bar{b}_1,\quad 3)\, c_{24}a_1=\bar{c}_{24}\bar{a}_1, \quad 4)\, c_{13}b_1=\bar{c}_{13}\bar{b}_1.
\end{equation}
The reduced system is
\begin{align}
&\mu v_{1,t}+6[c_{12}v_1v_2+c_{13}b_1|v_1|^2+c_{14}a_1v_1\bar{v}_2+\bar{c}_{14}\bar{a}_1\bar{v}_1v_2+c_{24}a_1|v_2|^2+
\bar{c}_{12}\bar{v}_1\bar{v}_2]v_{1,x}+v_{1,xxx}=0,\nonumber\\
&\mu v_{2,t}+6[c_{12}v_1v_2+c_{13}b_1|v_1|^2+c_{14}a_1v_1\bar{v}_2+\bar{c}_{14}\bar{a}_1\bar{v}_1v_2+c_{24}a_1|v_2|^2+
\bar{c}_{12}\bar{v}_1\bar{v}_2]v_{2,x}+v_{2,xxx}=0,\label{localiiN=4}
\end{align}
where the conditions (\ref{N=4localiicond}) are satisfied and $a\in \mathbb{R}$.

\noindent If we use the Type 1 approach by checking the equalities
\begin{align}
\displaystyle
&v_4=a_1\bar{v}_2(t,x) \Rightarrow \frac{g_4}{f}=a_1\frac{\bar{g}_2}{\bar{f}} \Rightarrow g_4=a_1\bar{g}_2, f=\bar{f},\\
&v_3=b_1\bar{v}_1(t,x) \Rightarrow \frac{g_3}{f}=b_1\frac{\bar{g}_1}{\bar{f}} \Rightarrow g_3=b_1\bar{g}_1, f=\bar{f},
\end{align}
we obtain the conditions on the parameters of one-soliton solution of the system (\ref{localiiN=4}) as
\begin{equation}
1)\, k_4=\bar{k}_2,\quad 2)\, k_3=\bar{k}_1,\quad 3)\, e^{\delta_4}=a_1e^{\bar{\delta}_2},\quad 4)\, e^{\delta_3}=b_1e^{\bar{\delta}_1},
\end{equation}
besides the conditions (\ref{N=4localiicond}). With all of these constraints one-soliton solution of the system (\ref{localiiN=4})
becomes
\begin{equation}\displaystyle
v_1=\frac{g_1}{f},\quad v_2=\frac{g_2}{f}
\end{equation}
where
\begin{align}
&g_1=e^{k_1x-\frac{k_1^3}{\mu}t
+\delta_1}+\frac{(k_1-k_2)(k_1-\bar{k}_1)(k_2-\bar{k}_1)}{(k_1+k_2)
(k_1+\bar{k}_1)(k_2+\bar{k}_1)}\frac{\bar{c}_{14}\bar{a}_1}
{(k_2^2-\bar{k}_1^2)}e^{(k_1+k_2+\bar{k}_1)x-\frac{k_1^3+k_2^3+\bar{k}_1^3}{\mu}t+\delta_1+\delta_2+\bar{\delta}_1}\nonumber\\
&+\frac{(k_1-k_2)(k_1-\bar{k}_2)(k_2-\bar{k}_2)}{(k_1+k_2)
(k_1+\bar{k}_2)(k_2+\bar{k}_2)}\frac{c_{24}a_1}
{(k_2^2-\bar{k}_2^2)}e^{(k_1+k_2+\bar{k}_2)x-\frac{k_1^3+k_2^3+\bar{k}_2^3}{\mu}t+\delta_1+\delta_2+\bar{\delta}_2}\nonumber\\
&+\frac{(k_1-\bar{k}_1)(k_1-\bar{k}_2)(\bar{k}_1-\bar{k}_2)}{(k_1+\bar{k}_1)
(k_1+\bar{k}_2)(\bar{k}_1+\bar{k}_2)}\frac{\bar{c}_{12}}
{(\bar{k}_1^2-\bar{k}_2^2)}e^{(k_1+\bar{k}_1+\bar{k}_2)x-\frac{k_1^3+\bar{k}_1^3+\bar{k}_2^3}{\mu}t+\delta_1+\bar{\delta}_1+\bar{\delta}_2},
\end{align}
\begin{align}
&g_2=e^{k_2x-\frac{k_2^3}{\mu}t
+\delta_2}-\frac{(k_1-k_2)(k_1-\bar{k}_1)(k_2-\bar{k}_1)}{(k_1+k_2)
(k_1+\bar{k}_1)(k_2+\bar{k}_1)}\frac{c_{13}b_1}
{(k_1^2-\bar{k}_1^2)}e^{(k_1+k_2+\bar{k}_1)x-\frac{k_1^3+k_2^3+\bar{k}_1^3}{\mu}t+\delta_1+\delta_2+\bar{\delta}_1}\nonumber\\
&-\frac{(k_1-k_2)(k_1-\bar{k}_2)(k_2-\bar{k}_2)}{(k_1+k_2)
(k_1+\bar{k}_2)(k_2+\bar{k}_2)}\frac{c_{14}a_1}
{(k_1^2-\bar{k}_2^2)}e^{(k_1+k_2+\bar{k}_2)x-\frac{k_1^3+k_2^3+\bar{k}_2^3}{\mu}t+\delta_1+\delta_2+\bar{\delta}_2}\nonumber\\
&+\frac{(k_2-\bar{k}_1)(k_2-\bar{k}_2)(\bar{k}_1-\bar{k}_2)}{(k_2+\bar{k}_1)
(k_2+\bar{k}_2)(\bar{k}_1+\bar{k}_2)}\frac{\bar{c}_{12}}
{(\bar{k}_1^2-\bar{k}_2^2)}e^{(k_2+\bar{k}_1+\bar{k}_2)x-\frac{k_2^3+\bar{k}_1^3+\bar{k}_2^3}{\mu}t+\delta_2+\bar{\delta}_1+\bar{\delta}_2},
\end{align}
and
\begin{align}
&f=1+\frac{c_{12}}{(k_1+k_2)^2}e^{(k_1+k_2)x-\frac{(k_1^3+k_2^3)}{\mu}t+\delta_1+\delta_2}
+\frac{\bar{a}_1\bar{c}_{14}}{(k_2+\bar{k}_1)^2}e^{(k_2+\bar{k}_1)x-\frac{(k_2^3+\bar{k}_1^3)}{\mu}t+\delta_2+\bar{\delta}_1}\nonumber\\
&+\frac{b_1c_{13}}{(k_1+\bar{k}_1)^2}e^{(k_1+\bar{k}_1)x-\frac{(k_1^3+\bar{k}_1^3)}{\mu}t+\delta_1+\bar{\delta}_1}
+\frac{a_1c_{24}}{(k_2+\bar{k}_2)^2}e^{(k_2+\bar{k}_2)x-\frac{(k_2^3+\bar{k}_2^3)}{\mu}t+\delta_2+\bar{\delta}_2}\nonumber\\
&+\frac{a_1c_{14}}{(k_1+\bar{k}_2)^2}e^{(k_1+\bar{k}_2)x-\frac{(k_1^3+\bar{k}_2^3)}{\mu}t+\delta_1+\bar{\delta}_2}
+\frac{\bar{c}_{12}}{(\bar{k}_1+\bar{k}_2)^2}e^{(\bar{k}_1+\bar{k}_2)x-\frac{(\bar{k}_1^3+\bar{k}_2^3)}{\mu}t+\bar{\delta}_1+\bar{\delta}_2}\nonumber\\
&+a_1b_1\frac{(k_1-k_2)(k_1-\bar{k}_1)(k_1-\bar{k}_2)(k_2-\bar{k}_1)(k_2-\bar{k}_2)(\bar{k}_1-\bar{k}_2)}
{(k_1+k_2)(k_1+\bar{k}_1)(k_1+\bar{k}_2)(k_2+\bar{k}_1)(k_2+\bar{k}_2)(\bar{k}_1+\bar{k}_2)}[\frac{c_{12}\bar{c}_{12}}
{a_1b_1(k_1^2-k_2^2)(\bar{k}_1^2-\bar{k}_2^2)}\nonumber\\
&-\frac{c_{13}c_{24}}{(k_1^2-\bar{k}_1^2)(k_2^2-\bar{k}_2^2)}+\frac{c_{14}c_{23}}{(k_1^2-\bar{k}_2^2)(k_2^2-\bar{k}_1^2)}]
e^{(k_1+k_2+\bar{k}_1+\bar{k}_2)x-\frac{k_1^3+k_2^3+\bar{k}_1^3+\bar{k}_2^3}{\mu}t+\delta_1+\delta_2+\bar{\delta}_1+\bar{\delta}_2}.
\end{align}

\noindent When the parameters $k_1, k_2$ are real then the Example 1 given in part i. is also valid for this case.

\subsection{Nonlocal Reductions for N=4}

Here we present two different nonlocal reductions of the coupled HI-mKdV system (\ref{coupledmKdVN}) for $N=4$.\\

\noindent \textbf{i.} $v_4=a_1v_2^{\varepsilon}$, $v_3=b_1v_1^{\varepsilon}$, $v_i^{\varepsilon}=v_i(\varepsilon_1 t,\varepsilon_2 x)$, $\varepsilon_i^2=1$, $i=1, 2$,  $a_1$ and $b_1$ are constants. We have a consistent reduction if the following conditions hold:
\begin{equation}
\varepsilon_1\varepsilon_2=1, \quad \rho=\rho^{\varepsilon}.
\end{equation}
To have a nonlocal system we have only one choice $(\varepsilon_1,\varepsilon_2)=(-1,-1)$. The relation $\rho=\rho^{\varepsilon}$ is satisfied if
\begin{equation}\label{N=4nonlocalcondi}
1)\, c_{12}=c_{34}a_1b_1,\quad 2)\, c_{14}a_1=c_{23}b_1.
\end{equation}
Then the reduced system is
\begin{align}
&\mu v_{1,t}(t,x)+6[c_{12}v_1(t,x)v_{2}(t,x)+c_{13}b_1v_1(t,x)v_1(-t,-x)+c_{23}b_1v_1(t,x)v_2(-t,-x)\nonumber\\
&+c_{23}b_1v_2(t,x)v_1(-t,-x)+c_{24}a_1v_2(t,x)v_2(-t,-x)+c_{12}v_1(-t,-x)v_2(-t,-x)]v_{1,x}(t,x)\nonumber\\
&+v_{1,xxx}(t,x)=0,\nonumber\\
 &\mu v_{2,t}(t,x)+6[c_{12}v_1(t,x)v_{2}(t,x)+c_{13}b_1v_1(t,x)v_1(-t,-x)+c_{23}b_1v_1(t,x)v_2(-t,-x)\nonumber\\
&+c_{23}b_1v_2(t,x)v_1(-t,-x)+c_{24}a_1v_2(t,x)v_2(-t,-x)+c_{12}v_1(-t,-x)v_2(-t,-x)]v_{2,x}(t,x)\nonumber\\
&+v_{2,xxx}(t,x)=0.\label{nonlocalreducediN=4}
\end{align}

\noindent To obtain one-soliton solution of the system (\ref{nonlocalreducediN=4}), we use the Type 2 approach since Type 1 gives trivial solution. Hence we take
\begin{align}
&g_4f^{\varepsilon}-a_1g_2^{\varepsilon}=0,\label{relationnonlocalia}\\
&g_3f^{\varepsilon}-b_1g_1^{\varepsilon}=0,\label{relationnonlocalib}
\end{align}
to get constraints for the parameters of one-soliton solution of the system (\ref{nonlocalreducediN=4}). Here we obtain the following cases satisfying both (\ref{relationnonlocalia}) and (\ref{relationnonlocalib}).\\

\noindent \textbf{Case A.} One of the sets of the relations solving the equation (\ref{relationnonlocalia}) is
\begin{equation}\label{caseA}
1)\, c_{12}=c_{13}=c_{23}=0, \quad 2)\, e^{\delta_2}=\sigma_1\frac{(k_2+k_4)}{\sqrt{a_1c_{24}}},\quad 3)\, e^{\delta_4}=\sigma_2\sqrt{a_1}\frac{(k_2+k_4)}{\sqrt{c_{24}}}, \sigma_i=\pm 1, i=1, 2.
\end{equation}
In this case the system (\ref{nonlocalreducediN=4}) becomes
\begin{align}
&\mu v_{1,t}(t,x)+6c_{24}a_1v_2(t,x)v_2(-t,-x)v_{1,x}(t,x)+v_{1,xxx}(t,x)=0,\nonumber\\
 &\mu v_{2,t}(t,x)+6c_{24}a_1v_2(t,x)v_2(-t,-x)v_{2,x}(t,x)+v_{2,xxx}(t,x)=0.\label{nonlocalreducediAN=4}
\end{align}

\noindent From the equation (\ref{relationnonlocalib}) we get the following cases:\\

\noindent \textbf{Case A.I.} In addition to the constraints (\ref{caseA}), one of the sets of the conditions that the equation (\ref{relationnonlocalib}) yields, is $\{ k_1=k_2, k_3=k_4, e^{\delta_3}=\sigma_1\sigma_2b_1e^{\delta_1}\}$ for $\sigma_i=\pm 1, i=1, 2$. Hence one-soliton solution of the system (\ref{nonlocalreducediAN=4}) is
\begin{equation}\displaystyle
v_1=\frac{e^{k_2x-\frac{k_2^3}{a}t+\delta_1}}{1+\sigma_1\sigma_2e^{(k_2+k_4)x-\frac{(k_2^3+k_4^3)}{a}t}},\,\,
v_2=\frac{\sigma_1(k_2+k_4)e^{k_2x-\frac{k_2^3}{a}t}}{\sqrt{a_1c_{24}}[1+\sigma_1\sigma_2e^{(k_2+k_4)x-\frac{(k_2^3+k_4^3)}{a}t}]}.
\end{equation}
\noindent Clearly, here we have $v_2=\xi v_1$, where $\xi=\frac{\sigma_1(k_2+k_4)}{e^{\delta_1}\sqrt{a_1c_{24}}}$. Hence the system
(\ref{nonlocalreducediAN=4}) reduces to a single nonlocal ST-symmetric mKdV equation \cite{GurPek2}
\begin{equation}
\mu v_{1,t}(t,x)+6c_{24}a_1\xi^2v_1(t,x)v_1(-t,-x)v_{1,x}(t,x)+v_{1,xxx}(t,x)=0.
\end{equation}
Note that another set of the conditions $\{ k_1=k_4, k_2=k_3, e^{\delta_3}=\sigma_1\sigma_2b_1e^{\delta_1}\}$ yielded form (\ref{relationnonlocalib}) gives similar solution.
\vspace{0.3cm}

\noindent \textbf{Example 2.} If we take the parameters as $k_4=2, \mu=k_2=e^{\delta_1}=a_1=c_{24}=\sigma_1=\sigma_2=1$, then one-soliton solution becomes
\begin{equation}\displaystyle
v_1=\frac{e^{x-t}}{1+e^{3x-9t}},
\end{equation}
and $v_2=3v_1$. This solution is finite and bounded for any $(x,t)$. The graph of the solution $v_1$ is given in Figure 2.
\begin{center}
\begin{figure}[h]
\begin{minipage}{1\textwidth}
\centering
\includegraphics[angle=0,scale=.18]{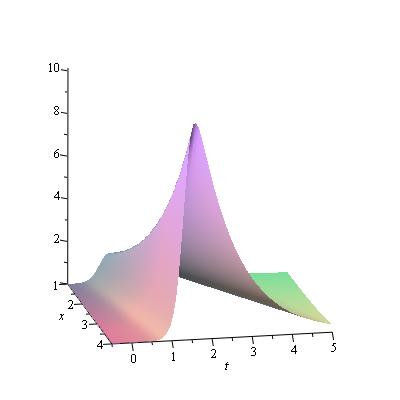}
\caption{One-soliton solution for (\ref{nonlocalreducediAN=4}) with the parameters $k_4=2, \mu=k_2=e^{\delta_1}=a_1=c_{24}=\sigma_1=\sigma_2=1$.}
\end{minipage}
\end{figure}
\end{center}

\noindent \textbf{Case A.II.} In addition to the constraints (\ref{caseA}), another set of the conditions obtained from the equation (\ref{relationnonlocalib}) is $\displaystyle \{ k_1=-k_3, k_2=-\frac{k_3^2}{k_4}, e^{\delta_3}=-b_1e^{\delta_1}\}$. In this case one-soliton solution of the system (\ref{nonlocalreducediAN=4}) is
\begin{align}\displaystyle
&v_1=\frac{e^{-k_3x+\frac{k_3^3}{\mu}t+\delta_1}-\sigma_1\sigma_2e^{(k_4-k_3-\frac{k_3^2}{k_4})x-\frac{1}{\mu}(k_4^3-k_3^3-\frac{k_3^6}{k_4^3})t+\delta_1}   }{1+\sigma_1\sigma_2e^{(k_4-\frac{k_3^2}{k_4})x-\frac{(k_4^3-\frac{k_3^6}{k_4^3})}{\mu}t}},\\
&v_2=\frac{\sigma_1(k_4-\frac{k_3^2}{k_4})e^{-\frac{k_3^2}{k_4}x+\frac{k_3^6}{\mu k_4^3}t}   }{\sqrt{a_1c_{24}}[1+\sigma_1\sigma_2e^{(k_4-\frac{k_3^2}{k_4})x-\frac{(k_4^3-\frac{k_3^6}{k_4^3})}{\mu}t}]},
\end{align}
for $\sigma_i=\pm 1, i=1, 2$.\\

\noindent \textbf{Example 3.} Take the parameters as $k_4=4, \mu=10, k_3=e^{\delta_1}=a_1=c_{24}=\sigma_1=\sigma_2=1$. Therefore one-soliton solution of
(\ref{nonlocalreducediAN=4}) becomes
\begin{equation}\displaystyle
v_1=\frac{e^{-x+\frac{1}{10}t}-e^{\frac{11}{4}x-\frac{4031}{640}t}}{1+e^{\frac{15}{4}x-\frac{819}{128}t}},
\quad v_2=\frac{15e^{-\frac{1}{4}x+\frac{1}{640}t}}{4(1+e^{\frac{15}{4}x-\frac{819}{128}t})}.
\end{equation}
The solutions are finite but not bounded. The graphs of the functions $v_1$ and $v_2$ are given in Figure 3(a) and Figure 3(b) respectively.
\begin{figure}[h!]
\centering     
\subfigure[]{\label{fig:locali1a}\includegraphics[width=28mm]{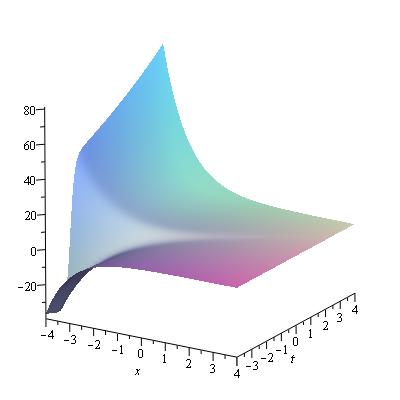}}\hspace{5cm}
\subfigure[]{\label{fig:locali1a}\includegraphics[width=28mm]{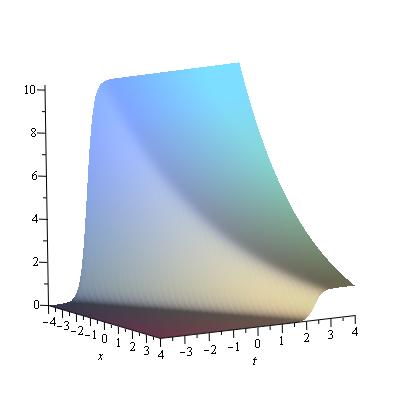}}
\caption{One-soliton solution of the system (\ref{nonlocalreducediAN=4}) with the parameters $k_4=4, \mu=10, k_3=e^{\delta_1}=a_1=c_{24}=\sigma_1=\sigma_2=1$.}
\end{figure}

\noindent \textbf{Case A.III.} In addition to the constraints (\ref{caseA}), the last set of the conditions obtained from the equation (\ref{relationnonlocalib}) is $\displaystyle \{ k_1=-k_3, e^{\delta_3}=-b_1e^{\delta_1}\frac{(k_3+k_4)(k_2+k_3)}{(k_3-k_4)(k_2-k_3)}\}$. Therefore one-soliton solution of the system (\ref{nonlocalreducediAN=4}) becomes
\begin{align}\displaystyle
&v_1=\frac{e^{-k_3x+\frac{k_3^3}{\mu}t+\delta_1}-\sigma_1\sigma_2\frac{(k_3+k_4)(k_2+k_3)}{(k_3-k_4)(k_2-k_3)}e^{(k_2-k_3+k_4)x-\frac{(k_2^3-k_3^3+k_4^3)}{\mu}t+\delta_1}   }{1+\sigma_1\sigma_2e^{(k_2+k_4)x-\frac{(k_2^3+k_4^3)}{\mu}t}},\\
&v_2=\frac{\sigma_1(k_2+k_4)e^{k_2x-\frac{k_2^3}{\mu}t}   }{\sqrt{a_1c_{24}}
[1+\sigma_1\sigma_2e^{(k_2+k_4)x-\frac{(k_2^3+k_4^3)}{\mu}t}]},
\end{align}
for $\sigma_i=\pm 1, i=1, 2$.
\vspace{0.3cm}

\noindent \textbf{Example 4.} If we choose the parameters as $k_2=2, k_3=-1, k_4=3, e^{\delta_1}=\frac{1}{10}, a_1=1, \mu=c_{24}=\sigma_1=\sigma_2=-1$ one-soliton solution of
(\ref{nonlocalreducediAN=4}) becomes
\begin{equation}\displaystyle
v_1=\frac{6e^{x+t}+e^{6x+36t}}{60(1+e^{5x+35t})},\quad |v_2|^2=\frac{25e^{4x+16t}}{(1+e^{5x+35t})^2}.
\end{equation}
Both of the solutions $v_1$ and $v_2$ are finite. The solution $v_2$ is bounded but $v_1$  not. The graphs of the functions $v_1$ and $|v_2|^2$ are given in Figure 4(a) and Figure 4(b) respectively.
\begin{figure}[h!]
\centering     
\subfigure[]{\label{fig:locali1a}\includegraphics[width=28mm]{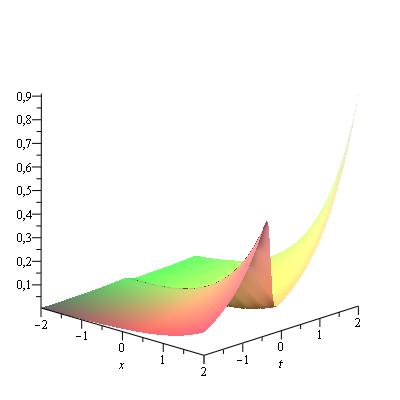}}\hspace{5cm}
\subfigure[]{\label{fig:locali1a}\includegraphics[width=28mm]{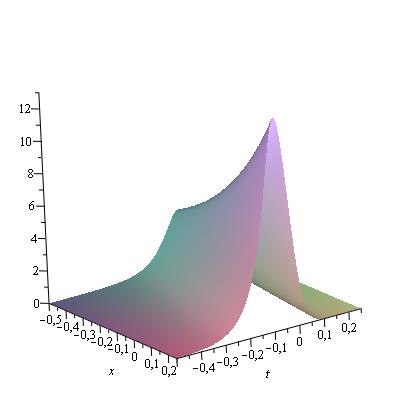}}
\caption{One-soliton solution of the system (\ref{nonlocalreducediAN=4}) with the parameters $k_2=2, k_3=-1, k_4=3, e^{\delta_1}=\frac{1}{10}, a_1=1, \mu=c_{24}=\sigma_1=\sigma_2=-1$.}
\end{figure}

\begin{rem}
 Note that if we start with the conditions $c_{12}=c_{23}=c_{24}=0$ then the system (\ref{nonlocalreducediN=4}) turns to be
\begin{align}
&\mu v_{1,t}(t,x)+6c_{13}b_1v_1(t,x)v_1(-t,-x)v_{1,x}(t,x)+v_{1,xxx}(t,x)=0,\nonumber\\
 &\mu v_{2,t}(t,x)+6c_{13}b_1v_1(t,x)v_1(-t,-x)v_{2,x}(t,x)+v_{2,xxx}(t,x)=0,\label{nonlocalreducediAsimilar}
\end{align}
which is very similar to the Case A.
\end{rem}

\vspace{0.3cm}
\noindent \textbf{Case B.} Another set of the constraints solving both of the equations (\ref{relationnonlocalia}) and (\ref{relationnonlocalib})  is
\begin{align}\label{caseB}\displaystyle
&1)\, c_{12}=c_{23}=k_2=0,\quad 2)\,k_1=\frac{k_4[a_1k_4^2(k_3+k_4)+e^{2\delta_4}c_{24}(k_3-k_4)  ]}{-k_4^2a_1(k_3+k_4)+e^{2\delta_4}c_{24}(k_3-k_4)},\nonumber\\
& 3)\, e^{\delta_2}=\sigma_1\frac{k_4}{\sqrt{a_1c_{24}}},\quad 4)\, e^{\delta_3}=\sigma_2\frac{\sqrt{b_1(k_3^2-k_4^2)}}{\sqrt{c_{13}}}\frac{(e^{2\delta_4}c_{24}-a_1k_4^2)(k_3+k_4)}{a_1k_4^2(k_3+k_4)
-e^{2\delta_4}c_{24}(k_3-k_4)},\nonumber\\
&5)\,  e^{\delta_1}=\frac{\sigma_1e^{\delta_4}c_{24}(k_3-k_4)^2(k_3+k_4)[a_1k_4^2-e^{2\delta_4}c_{24}]}
{\sigma_2[a_1k_4^2(k_3+k_4)-e^{2\delta_4}c_{24}(k_3-k_4)]k_4\sqrt{a_1b_1c_{13}c_{24}(k_3^2-k_4^2)}},
\end{align}
where $\sigma_i=\pm 1, i=1, 2$. In this case the system (\ref{nonlocalreducediN=4}) reduces to
\begin{align}
&\mu v_{1,t}(t,x)+6(c_{13}b_1v_1(t,x)v_1(-t,-x)+c_{24}a_1v_2(t,x)v_2(-t,-x))v_{1,x}(t,x)+v_{1,xxx}(t,x)=0,\nonumber\\
 &\mu v_{2,t}(t,x)+6(c_{13}b_1v_1(t,x)v_1(-t,-x)+c_{24}a_1v_2(t,x)v_2(-t,-x))v_{2,x}(t,x)+v_{2,xxx}(t,x)=0.\label{nonlocalreducediBN=4}
\end{align}
One-soliton solution of the reduced system (\ref{nonlocalreducediBN=4}) is given by
\begin{equation}\displaystyle
v_1=\frac{g_1}{f}, \quad v_2=\frac{g_2}{f},
\end{equation}
where
\begin{align}
g_1&=e^{k_1x-\frac{k_1^3}{\mu}t+\delta_1}+\frac{(k_1-k_4)c_{24}}{(k_1+k_4)k_4^2}e^{(k_1+k_4)x-\frac{(k_1^3+k_4^3)}{\mu}t+\delta_1+\delta_2+\delta_4},\\
g_2&=e^{\delta_2}+\frac{c_{13}}{(k_1+k_3)^2}e^{(k_1+k_3)x-\frac{(k_1^3+k_3^3)}{\mu}t+\delta_1+\delta_2+\delta_3},\\
f&=1+\frac{c_{13}}{(k_1+k_3)^2}e^{(k_1+k_3)x-\frac{(k_1^3+k_3^3)}{\mu}t+\delta_1+\delta_3}
+\frac{c_{24}}{k_4^2}e^{k_4x-\frac{k_4^3}{\mu}t+\delta_2+\delta_4}\nonumber\\
&+\frac{(k_1-k_4)(k_3-k_4)c_{13}c_{24}}{(k_1+k_4)(k_3+k_4)(k_1+k_3)^2k_4^2}e^{(k_1+k_3+k_4)x-\frac{(k_1^3+k_3^3+k_4^3)}{\mu}t+\delta_1+\delta_2+\delta_3+\delta_4}.
\end{align}

\vspace{0.5cm}

\noindent \textbf{Example 5.} Let us take the parameters as $k_3=2, a=4, k_4=a_1=b_1=c_{13}=c_{24}=\sigma_1=\sigma_2=1, e^{\delta_4}=2$. In this case one-soliton solution of
(\ref{nonlocalreducediBN=4}) becomes
\begin{equation}\displaystyle
v_1=\frac{-3\sqrt{3}[2e^{7x-\frac{343}{4}t}+3e^{8x-86t}]}{1+2e^{9x-\frac{351}{4}t}+2e^{x-\frac{1}{4}t}+e^{10x-88t}},\quad v_2=\frac{1+2e^{9x-\frac{351}{4}t}}{1+2e^{9x-\frac{351}{4}t}+2e^{x-\frac{1}{4}t}+e^{10x-88t}}.
\end{equation}
These solutions are finite and bounded. The graphs of the functions $v_1$ and $v_2$ are given in Figure 5(a) and Figure 5(b) respectively.
\begin{figure}[h!]
\centering     
\subfigure[]{\label{fig:locali1a}\includegraphics[width=28mm]{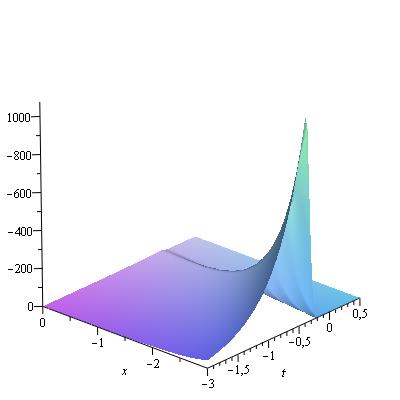}}\hspace{5cm}
\subfigure[]{\label{fig:locali1a}\includegraphics[width=28mm]{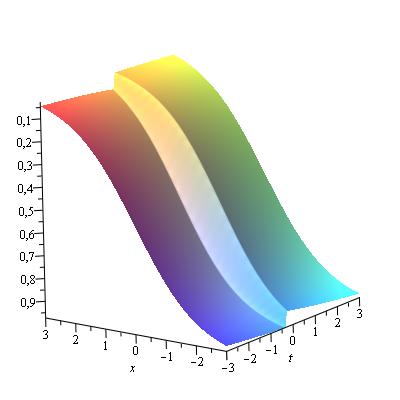}}
\caption{One-soliton solution of the system (\ref{nonlocalreducediBN=4}) with the parameters $k_3=2, \mu=4, k_4=a_1=b_1=c_{13}=c_{24}=\sigma_1=\sigma_2=1, e^{\delta_4}=2$.}
\end{figure}
\vspace{0.3cm}

\noindent \textbf{Case C.} The last set of the constraints solving both of the equations (\ref{relationnonlocalia}) and (\ref{relationnonlocalib})  is
\begin{align}\label{caseC}\displaystyle
&1)\,c_{12}=c_{23}=0,\quad 2)\, e^{\delta_1}=\sigma_1(k_1+k_3)\sqrt{\frac{(k_1+k_2)(k_1+k_4)}{(k_1-k_2)(k_4-k_1)b_1c_{13}}},\nonumber\\
&3)\, e^{\delta_2}=\sigma_2(k_2+k_4)\sqrt{\frac{(k_1+k_2)(k_2+k_3)}{(k_1-k_2)(k_2-k_3)a_1c_{24}}},\quad
4)\, e^{\delta_3}=\sigma_3(k_1+k_3)\sqrt{b_1\frac{(k_2+k_3)(k_3+k_4)}{(k_2-k_3)(k_4-k_3)c_{13}}},\nonumber\\
&5)\, e^{\delta_4}=\sigma_4(k_2+k_4)\sqrt{a_1\frac{(k_1+k_4)(k_3+k_4)}{(k_1-k_4)(k_3-k_4)c_{24}}},\, \sigma_i=\pm 1, i=1,2,3,4.
\end{align}
Here the reduced system is the same with (\ref{nonlocalreducediBN=4}). One-soliton solution of this system is given by
\begin{equation}\displaystyle
v_1=\frac{g_1}{f}, \quad v_2=\frac{g_2}{f},
\end{equation}
where
\begin{align}
g_1&=e^{k_1x-\frac{k_1^3}{\mu}t+\delta_1}+\frac{(k_1-k_2)(k_1-k_4)c_{24})}{(k_1+k_2)(k_1+k_4)(k_2+k_4)^2}e^{(k_1+k_2+k_4)x
-\frac{(k_1^3+k_2^3+k_4^3)}{\mu}t+\delta_1+\delta_2+\delta_4},\\
g_2&=e^{k_2x-\frac{k_2^3}{\mu}t+\delta_2}-\frac{(k_1-k_2)(k_2-k_3)c_{13}}{(k_1+k_2)(k_2+k_3)(k_1+k_3)^2}e^{(k_1+k_2+k_3)x-\frac{(k_1^3+k_2^3+k_3^3)}{\mu}t
+\delta_1+\delta_2+\delta_3},\\
f&=1+\frac{c_{13}}{(k_1+k_3)^2}e^{(k_1+k_3)x-\frac{(k_1^3+k_3^3)}{\mu}t+\delta_1+\delta_3}
+\frac{c_{24}}{(k_2+k_4)^2}e^{(k_2+k_4)x-\frac{(k_2^3+k_4^3)}{\mu}t+\delta_2+\delta_4}\nonumber\\
&-\frac{(k_1-k_2)(k_1-k_4)(k_2-k_3)(k_3-k_4)c_{13}c_{24}e^{(k_1+k_2+k_3+k_4)x-\frac{(k_1^3+k_2^3+k_3^3+k_4^3)}
{\mu}t+\delta_1+\delta_2+\delta_3+\delta_4}}{(k_1+k_2)(k_1+k_4)(k_2+k_3)(k_3+k_4)(k_1+k_3)^2(k_2+k_4)^2}.
\end{align}
\vspace{0.5cm}

\noindent \textbf{Example 6.} Let us choose the parameters as $k_1=4, k_2=1, k_3=3, k_4=2, \mu=3, a_1=b_1=c_{13}=c_{24}=\sigma_i=1, i=1, 2, 3, 4$. Hence we obtain the following one-soliton solution of (\ref{nonlocalreducediBN=4}):
\begin{equation}\displaystyle
|v_1|^2=\frac{W_1}{Y}, \quad |v_2|^2=\frac{W_2}{Y^2},
\end{equation}
where
\begin{align}
W_1&=245(2e^{6x}+e^{6t})e^{8x+18t},\\
W_2&=15e^{2x+\frac{14}{3}t}[882e^{24x+34t}+196e^{10x+\frac{284}{3}t}+9e^{34x+\frac{2}{3}t}+109e^{14x+\frac{202}{3}t}\nonumber\\
&+452e^{20x+\frac{184}{3}t}+100e^{6x+122t}+450e^{28x+\frac{20}{3}t}+2e^{128t}],\\
Y&=e^{\frac{200}{3}t}+e^{20x}+98e^{10x+\frac{100}{3}t}+50e^{14x+6t}+50e^{6x+\frac{182}{3}t}.
\end{align}
These solutions are finite and bounded. The graphs of the functions $|v_1|^2$ and $|v_2|^2$ are given in Figure 6(a) and Figure 6(b) respectively.
\begin{figure}[h!]
\centering     
\subfigure[]{\label{fig:locali1a}\includegraphics[width=28mm]{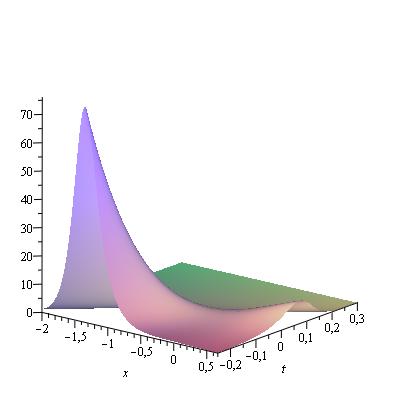}}\hspace{5cm}
\subfigure[]{\label{fig:locali1a}\includegraphics[width=28mm]{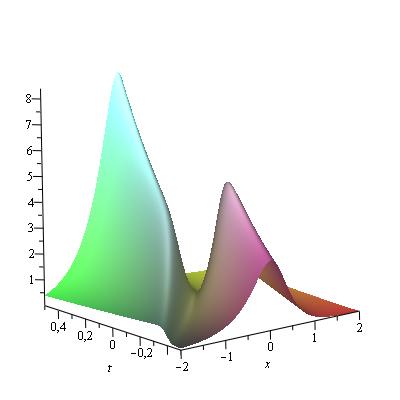}}
\caption{One-soliton solution of the system (\ref{nonlocalreducediBN=4}) with the parameters $k_1=4, k_2=1, k_3=3, k_4=2, \mu=3, a_1=b_1=c_{13}=c_{24}=\sigma_i=1, i=1, 2, 3, 4$.}
\end{figure}
\vspace{0.3cm}

\noindent \textbf{ii.} \textbf{$v_4=a_1\bar{v}_2^{\varepsilon}$, $v_3=b_1\bar{v}_1^{\varepsilon}$}, $ \bar{v}_i^{\varepsilon}=\bar{v}_i(\varepsilon_1 t,\varepsilon_2 x)$, $\varepsilon_i^2=1$, $i= 1, 2$, $a_1$ and $b_1$ are constants. To have a consistent reduction the following conditions must hold:
\begin{equation}
\mu=\bar{\mu}\varepsilon_1\varepsilon_2, \quad \rho=\bar{\rho}^{\varepsilon}.
\end{equation}
 The relation $\rho=\bar{\rho}^{\varepsilon}$ is satisfied if
\begin{equation}\label{N=4nonlocalcondii}
1)\, c_{12}=\bar{c}_{34}\bar{a}_1\bar{b}_1,\quad 2)\, c_{14}a_1=\bar{c}_{23}\bar{b}_1,\quad 3)\, c_{13}b_1=\bar{c}_{13}\bar{b}_1,\quad 4)\, c_{24}a_1=\bar{c}_{24}\bar{a}_1.
\end{equation}

\noindent Here $(\varepsilon_1,\varepsilon_2)=\{(-1,1), (1,-1), (-1, -1)\}$. Hence we have $T$-, $S$-, and $ST$-symmetric nonlocal coupled complex HI-mKdV systems.
If we use the Type 1 approach, the conditions that the parameters of the one-soliton solutions of nonlocal coupled complex HI-mKdV systems satisfy are obtained from
\begin{align}
&\frac{g_4}{f}=a_1\frac{\bar{g}_2^{\varepsilon}}{\bar{f}^{\varepsilon}} \Rightarrow g_4=a_1\bar{g}_2^{\varepsilon}, f=\bar{f}^{\varepsilon},\\
&\frac{g_3}{f}=b_1\frac{\bar{g}_1^{\varepsilon}}{\bar{f}^{\varepsilon}} \Rightarrow g_3=b_1\bar{g}_1^{\varepsilon}, f=\bar{f}^{\varepsilon}.
\end{align}
Hence we obtain the following constraints:
\begin{equation}
1)\,k_4=\varepsilon_2\bar{k}_2,\,\, 2)\, k_3=\varepsilon_2\bar{k}_1,\,\, 3)\, e^{\delta_4}=a_1e^{\bar{\delta}_2},\,\, 4)\, e^{\delta_3}=b_1e^{\bar{\delta}_1},
\end{equation}
with $\mu=\bar{\mu}\varepsilon_1\varepsilon_2$ and (\ref{N=4nonlocalcondii}). Hence one-soliton solutions of the nonlocal coupled complex HI-mKdV systems are
\begin{equation}
  \displaystyle v_1=\frac{g_1}{f}, \quad v_2=\frac{g_2}{f},
\end{equation}
where
\begin{align}
&g_1=e^{k_1x-\frac{k_1^3}{\mu}t+\delta_1}\nonumber\\
&+\frac{(k_1-k_2)}{(k_1+k_2)}\frac{(k_1-\varepsilon_2\bar{k}_1)}
{(k_1+\varepsilon_2\bar{k}_1)}\frac{(k_2-\varepsilon_2\bar{k}_1)}
{(k_2+\varepsilon_2\bar{k}_1)}\frac{c_{23}b_1}{(k_2^2-\bar{k}_1^2)}e^{(k_1+k_2+\varepsilon_2\bar{k}_1)x-\frac{(k_1^3+k_2^3+\varepsilon_2\bar{k}_1^3)}{\mu}t
+\delta_1+\delta_2+\bar{\delta}_1}\nonumber\\
&+\frac{(k_1-k_2)}{(k_1+k_2)}\frac{(k_1-\varepsilon_2\bar{k}_2)}
{(k_1+\varepsilon_2\bar{k}_2)}\frac{(k_2-\varepsilon_2\bar{k}_2)}
{(k_2+\varepsilon_2\bar{k}_2)}\frac{c_{24}a_1}{(k_2^2-\bar{k}_2^2)}e^{(k_1+k_2+\varepsilon_2\bar{k}_2)x-\frac{(k_1^3+k_2^3+\varepsilon_2\bar{k}_2^3)}{\mu}t
+\delta_1+\delta_2+\bar{\delta}_2}\nonumber\\
&+\frac{(k_1-\varepsilon_2\bar{k}_1)}{(k_1+\varepsilon_2\bar{k}_1)}\frac{(k_1-\varepsilon_2\bar{k}_2)}
{(k_1+\varepsilon_2\bar{k}_2)}\frac{(\bar{k}_1-\bar{k}_2)}
{(\bar{k}_1+\bar{k}_2)}\frac{\bar{c}_{12}}{(\bar{k}_1^2-\bar{k}_2^2)}e^{(k_1+\varepsilon_2\bar{k}_1+\varepsilon_2\bar{k}_2)x-\frac{(k_1^3+\varepsilon_2\bar{k}_1^3+\varepsilon_2\bar{k}_2^3)}
{\mu}t
+\delta_1+\bar{\delta}_1+\bar{\delta}_2},
\end{align}
\begin{align}
&g_2=e^{k_2x-\frac{k_2^3}{\mu}t+\delta_2}\nonumber\\
&-\frac{(k_1-k_2)}{(k_1+k_2)}\frac{(k_1-\varepsilon_2\bar{k}_1)}
{(k_1+\varepsilon_2\bar{k}_1)}\frac{(k_2-\varepsilon_2\bar{k}_1)}
{(k_2+\varepsilon_2\bar{k}_1)}\frac{c_{13}b_1}{(k_1^2-\bar{k}_1^2)}e^{(k_1+k_2+\varepsilon_2\bar{k}_1)x-\frac{(k_1^3+k_2^3+\varepsilon_2\bar{k}_1^3)}{\mu}t
+\delta_1+\delta_2+\bar{\delta}_1}\nonumber\\
&-\frac{(k_1-k_2)}{(k_1+k_2)}\frac{(k_1-\varepsilon_2\bar{k}_2)}
{(k_1+\varepsilon_2\bar{k}_2)}\frac{(k_2-\varepsilon_2\bar{k}_2)}
{(k_2+\varepsilon_2\bar{k}_2)}\frac{\bar{c}_{23}\bar{b}_1}{(k_1^2-\bar{k}_2^2)}e^{(k_1+k_2+\varepsilon_2\bar{k}_2)x-\frac{(k_1^3+k_2^3+\varepsilon_2\bar{k}_2^3)}{\mu}t
+\delta_1+\delta_2+\bar{\delta}_2}\nonumber\\
&+\frac{(k_2-\varepsilon_2\bar{k}_1)}{(k_2+\varepsilon_2\bar{k}_1)}\frac{(k_2-\varepsilon_2\bar{k}_2)}
{(k_2+\varepsilon_2\bar{k}_2)}\frac{(\bar{k}_1-\bar{k}_2)}
{(\bar{k}_1+\bar{k}_2)}\frac{\bar{c}_{12}}{(\bar{k}_1^2-\bar{k}_2^2)}e^{(k_2+\varepsilon_2\bar{k}_1+\varepsilon_2\bar{k}_2)x-\frac{(k_2^3+\varepsilon_2\bar{k}_1^3+\varepsilon_2\bar{k}_2^3)}{\mu}t
+\delta_2+\bar{\delta}_1+\bar{\delta}_2},
\end{align}
and
\begin{align}
&f=1+\frac{(k_1-k_2)}{(k_1+k_2)}\frac{c_{12}}{(k_1^2-k_2^2)}e^{(k_1+k_2)x-\frac{(k_1^3+k_2^3)}{\mu}t+\delta_1+\delta_2}\nonumber\\
&+\frac{(k_1-\varepsilon_2\bar{k}_1)}{(k_1+\varepsilon_2\bar{k}_1)}\frac{c_{13}b_1}{(k_1^2-\bar{k}_1^2)}e^{(k_1+\varepsilon_2\bar{k}_1)x
-\frac{(k_1^3+\varepsilon_2\bar{k}_1^3)}{\mu}t+\delta_1+\bar{\delta}_1}\nonumber\\
&+\frac{(k_1-\varepsilon_2\bar{k}_2)}{(k_1+\varepsilon_2\bar{k}_2)}\frac{\bar{c}_{23}\bar{b}_1}{(k_1^2-\bar{k}_2^2)}e^{(k_1+\varepsilon\bar{k}_2)x
-\frac{(k_1^3+\varepsilon_2\bar{k}_2^3)}{\mu}t+\delta_1+\bar{\delta}_2}+
\frac{(k_2-\varepsilon_2\bar{k}_1)}{(k_2+\varepsilon_2\bar{k}_1)}\frac{c_{23}b_1}{(k_2^2-\bar{k}_1^2)}e^{(k_2+\varepsilon\bar{k}_1)x
-\frac{(k_2^3+\varepsilon_2\bar{k}_1^3)}{\mu}t+\delta_2+\bar{\delta}_1}\nonumber\\
&+\frac{(k_2-\varepsilon_2\bar{k}_2)}{(k_2+\varepsilon_2\bar{k}_2)}\frac{c_{24}a_1}{(k_2^2-\bar{k}_2^2)}e^{(k_2+\varepsilon\bar{k}_2)x
-\frac{(k_2^3+\varepsilon_2\bar{k}_2^3)}{\mu}t+\delta_2+\bar{\delta}_2}+
\frac{(\bar{k}_1-\bar{k}_2)}{(\bar{k}_1+\bar{k}_2)}\frac{\bar{c}_{12}}{(\bar{k}_1^2-\bar{k}_2^2)}e^{\varepsilon_2(\bar{k}_1+\bar{k}_2)x
-\frac{\varepsilon_2(\bar{k}_1^3+\bar{k}_2^3)}{\mu}t+\bar{\delta}_1+\bar{\delta}_2}\nonumber\\
&+\frac{(k_1-k_2)}{(k_1+k_2)}\frac{(k_1-\varepsilon_2\bar{k}_1)}{(k_1+\varepsilon_2\bar{k}_1)}
\frac{(k_1-\varepsilon_2\bar{k}_2)}{(k_1+\varepsilon_2\bar{k}_2)}\frac{(k_2-\varepsilon_2\bar{k}_1)}{(k_2+\varepsilon_2\bar{k}_1)}
\frac{(k_2-\varepsilon_2\bar{k}_2)}{(k_2-\varepsilon_2\bar{k}_2)}\frac{(\bar{k}_1-\bar{k}_2)}{(\bar{k}_1+\bar{k}_2)}\Big[\frac{c_{12}}{(k_1^2-k_2^2)}
\frac{\bar{c}_{12}}{a_1b_1(\bar{k}_1^2-\bar{k}_2^2)}\nonumber\\
&-\frac{c_{13}}{(k_1^2-\bar{k}_1^2)}
\frac{c_{24}}{(k_2^2-\bar{k}_2^2)}+\frac{\bar{c}_{23}\bar{b}_1}{a_1(k_1^2-\bar{k}_2^2)}
\frac{c_{23}}{(k_2^2-\bar{k}_1^2)}\Big]a_1b_1e^{(k_1+k_2+\varepsilon_2\bar{k}_1+\varepsilon_2\bar{k}_2)x
-\frac{k_1^3+k_2^3+\varepsilon_2\bar{k}_1^3+\varepsilon_2\bar{k}_2^3}{\mu}t+\delta_1+\delta_2+\bar{\delta}_1+\bar{\delta}_2}
\end{align}

\vspace{0.3cm}
\noindent The reduced nonlocal complex HI-mKdV systems are given by,\\

\noindent a)\, \textbf{T-Symmetric Nonlocal Complex HI-mKdV System:}
\begin{align}\label{TsymmcompN=4sys}
&\mu v_{1,t}(t,x)+6[c_{12}v_1(t,x)v_{2}(t,x)+c_{13}b_1v_1(t,x)\bar{v}_1(-t,x)+\bar{c}_{23}\bar{b}_1v_1(t,x)\bar{v}_2(-t,x)\nonumber\\
&+c_{23}b_1v_2(t,x)\bar{v}_1(-t,x)+c_{24}a_1v_2(t,x)\bar{v}_2(-t,x)+\bar{c}_{12}\bar{v}_1(-t,x)\bar{v}_2(-t,x)]v_{1,x}(t,x)\nonumber\\
&+v_{1,xxx}(t,x)=0,\nonumber\\
 &\mu v_{2,t}(t,x)+6[c_{12}v_1(t,x)v_{2}(t,x)+c_{13}b_1v_1(t,x)\bar{v}_1(-t,x)+\bar{c}_{23}\bar{b}_1v_1(t,x)\bar{v}_2(-t,x)\nonumber\\
&+c_{23}b_1v_2(t,x)\bar{v}_1(-t,x)+c_{24}a_1v_2(t,x)\bar{v}_2(-t,x)+\bar{c}_{12}\bar{v}_1(-t,x)\bar{v}_2(-t,x)]v_{2,x}(t,x)\nonumber\\
&+v_{2,xxx}(t,x)=0.
\end{align}

\vspace{0.3cm}

\noindent \textbf{Example 7.} Consider the following parameters: $\mu=2i, a_1=b_1=1, k_1=1, k_2=2, c_{12}=c_{13}=c_{23}=c_{24}=e^{\delta_1}=e^{\delta_2}=1$.
Hence the pair of the solutions of (\ref{TsymmcompN=4sys}) becomes
\begin{equation}
\displaystyle
|v_1|^2=\frac{81e^{2x}}{(4e^{3x}+9\cos(\frac{9}{2}t))^2+81\sin^2(\frac{9}{2}t)},\,
|v_2|^2=\frac{81e^{4x}}{(4e^{3x}+9\cos(\frac{9}{2}t))^2+81\sin^2(\frac{9}{2}t)}.
\end{equation}
Both of the functions have singularity at $(t,x)=(\frac{(4n+2)}{9}\pi, x=\frac{2}{3}\ln (\frac{3}{2}))$. The graph of the solution
is given in Figure 7.

\begin{center}
\begin{figure}[h]
\begin{minipage}{1\textwidth}
\centering
\includegraphics[angle=0,scale=.18]{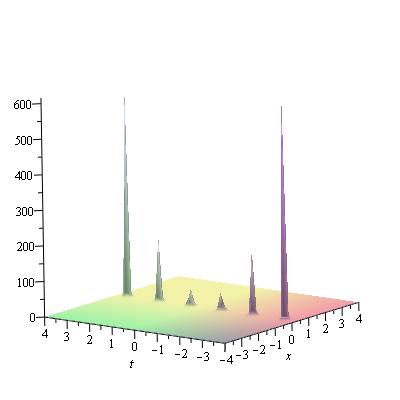}
\caption{Singular solution for (\ref{TsymmcompN=4sys}) with $\mu=2i, a_1=b_1=1, k_1=1, k_2=2, c_{12}=c_{13}=c_{23}=c_{24}=e^{\delta_1}=e^{\delta_2}=1$.}
\end{minipage}
\end{figure}
\end{center}

\newpage
\noindent b)\, \textbf{S-Symmetric Nonlocal Complex HI-mKdV System:}
\begin{align}\label{SsymmcompN=4sys}
&\mu v_{1,t}(t,x)+6[c_{12}v_1(t,x)v_{2}(t,x)+c_{13}b_1v_1(t,x)\bar{v}_1(t,-x)+\bar{c}_{23}\bar{b}_1v_1(t,x)\bar{v}_2(t,-x)\nonumber\\
&+c_{23}b_1v_2(t,x)\bar{v}_1(t,-x)+c_{24}a_1v_2(t,x)\bar{v}_2(t,-x)+\bar{c}_{12}\bar{v}_1(t,-x)\bar{v}_2(t,-x)]v_{1,x}(t,x)\nonumber\\
&+v_{1,xxx}(t,x)=0,\nonumber\\
 &\mu v_{2,t}(t,x)+6[c_{12}v_1(t,x)v_{2}(t,x)+c_{13}b_1v_1(t,x)\bar{v}_1(t,-x)+\bar{c}_{23}\bar{b}_1v_1(t,x)\bar{v}_2(t,-x)\nonumber\\
&+c_{23}b_1v_2(t,x)\bar{v}_1(t,-x)+c_{24}a_1v_2(t,x)\bar{v}_2(t,-x)+\bar{c}_{12}\bar{v}_1(t,-x)\bar{v}_2(t,-x)]v_{2,x}(t,x)\nonumber\\
&+v_{2,xxx}(t,x)=0.
\end{align}

\vspace{0.7cm}

\noindent \textbf{Example 8.} Take the following set of the parameters: $\mu=4i, a_1=b_1=1, k_1=\frac{i}{4}, k_2=\frac{i}{2}, c_{12}=c_{13}=c_{23}=c_{24}=e^{\delta_1}=e^{\delta_2}=1$. Hence the solutions of (\ref{SsymmcompN=4sys}) become
\begin{equation}
\displaystyle
|v_1|^2=\frac{81e^{\frac{1}{128}t}}{(64e^{\frac{9}{256}t}-9\cos(\frac{3}{4}x))^2+\sin^2(\frac{3}{4}x)}, \,|v_2|^2=\frac{81e^{\frac{1}{16}t}}{(64e^{\frac{9}{256}t}-9\cos(\frac{3}{4}x))^2+\sin^2(\frac{3}{4}x)}.
\end{equation}
The solutions are finite for $t\geq 0$ and bounded for any $(x,t)$. The graph of the function $|v_1|^2$ is given in Figure 8.
\begin{center}
\begin{figure}[h]
\begin{minipage}{1\textwidth}
\centering
\includegraphics[angle=0,scale=.18]{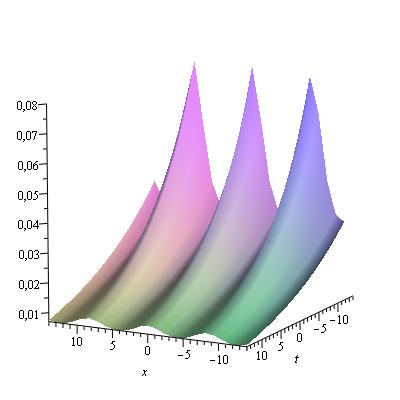}
\caption{Periodic wave solution for (\ref{SsymmcompN=4sys}) with $\mu=4i, a_1=b_1=1, k_1=\frac{i}{4}, k_2=\frac{i}{2}, c_{12}=c_{13}=c_{23}=c_{24}=e^{\delta_1}=e^{\delta_2}=1$.}
\end{minipage}
\end{figure}
\end{center}

\noindent c)\, \textbf{ST-Symmetric Nonlocal Complex HI-mKdV System:}
\begin{align}\label{STsymmcompN=4sys}
&\mu v_{1,t}(t,x)+6[c_{12}v_1(t,x)v_{2}(t,x)+c_{13}b_1v_1(t,x)\bar{v}_1(-t,-x)+\bar{c}_{23}\bar{b}_1v_1(t,x)\bar{v}_2(-t,-x)\nonumber\\
&+c_{23}b_1v_2(t,x)\bar{v}_1(-t,-x)+c_{24}a_1v_2(t,x)\bar{v}_2(-t,-x)+\bar{c}_{12}\bar{v}_1(-t,-x)\bar{v}_2(-t,-x)]v_{1,x}(t,x)\nonumber\\
&+v_{1,xxx}(t,x)=0,\nonumber\\
 &\mu v_{2,t}(t,x)+6[c_{12}v_1(t,x)v_{2}(t,x)+c_{13}b_1v_1(t,x)\bar{v}_1(-t,-x)+\bar{c}_{23}\bar{b}_1v_1(t,x)\bar{v}_2(-t,-x)\nonumber\\
&+c_{23}b_1v_2(t,x)\bar{v}_1(-t,-x)+c_{24}a_1v_2(t,x)\bar{v}_2(-t,-x)+\bar{c}_{12}\bar{v}_1(-t,-x)\bar{v}_2(-t,-x)]v_{2,x}(t,x)\nonumber\\
&+v_{2,xxx}(t,x)=0.
\end{align}
\vspace{0.5cm}
\noindent \textbf{Example 9.} Let us take the parameters as $\mu=20, a_1=1, b_1=2i, k_1=-\frac{i}{2}, k_2=2i, c_{12}=c_{13}=c_{23}=i, c_{24}=1, e^{\delta_1}=20, e^{\delta_2}=2i$. In this case the solutions of (\ref{STsymmcompN=4sys}) become
\begin{equation}
\displaystyle
|v_1|^2=\frac{32400}{102481+5760\cos(\frac{3}{2}x+\frac{63}{160}t)}, \,|v_2|^2=\frac{324}{102481+5760\cos(\frac{3}{2}x+\frac{63}{160}t)}.
\end{equation}
The above functions are finite and bounded for any $(x,t)$. The graph of the function $|v_1|^2$ is given in Figure 9.
\begin{center}
\begin{figure}[h]
\begin{minipage}{1\textwidth}
\centering
\includegraphics[angle=0,scale=.18]{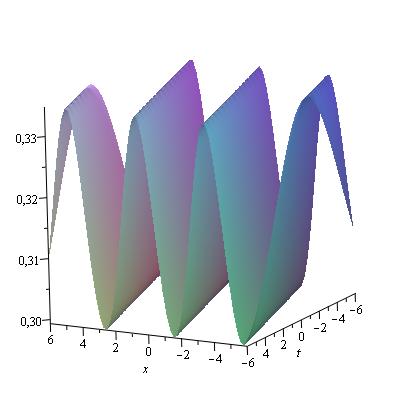}
\caption{Periodic wave solution for (\ref{STsymmcompN=4sys}) with $\mu=20, a_1=1, b_1=2i, k_1=-\frac{i}{2}, k_2=2i, c_{12}=c_{13}=c_{23}=i, c_{24}=1, e^{\delta_1}=20, e^{\delta_2}=2i$.}
\end{minipage}
\end{figure}
\end{center}

\section{Concluding Remarks}

In this work we studied a type of coupled HI-mKdV systems. We presented all possible consistent local and
Ablowitz-Musslimani type nonlocal reductions for general $N$. We obtained the Hirota bilinear forms of the systems and recalled
the multi-soliton solutions expressed by pfaffians for any $N$. We mentioned about local and nonlocal reductions of
the coupled HI-mKdV system for $N=2$ and one-soliton solution of the reduced systems which have been already presented
in our previous works. We examined all possible reductions for $N=3$. We noticed that the system for $N=3$ has reductions to systems of two equations but
if we use Type 1 approach to find one-soliton solution of these systems, because of the set of the constraints obtained, these systems of two equations reduce to single equations. We studied the case for $N=4$ in detail. We found one-soliton solution of $N=4$ coupled HI-mKdV system
by using the pfaffians. We have two local and two nonlocal consistent reductions for this system. These reductions yield local and nonlocal systems of two equations. By using the one-soliton solution of the
system with the reductions we also obtained one-soliton solutions of the local and nonlocal reduced systems by the help of Type 1 and Type 2 approaches.

\section{Appendix}

Here we present local and nonlocal reductions, and one-soliton solution of the system (\ref{coupledmKdVN}) for $N=3$. Note that
even this system has reductions to systems of two equations, if we require Type 1 one-soliton solution \cite{GurPek2}, all the systems of two equations reduce to single well-known equations.

\subsection{N=3 Coupled HI-mKdV System}

The system (\ref{coupledmKdVN}) for $N=3$ is
\begin{align}
& \mu v_{1,t}+3\rho v_{1,x}+v_{1,xxx}=0,\label{N=3a}\\
& \mu v_{2,t}+3\rho v_{2,x}+v_{2,xxx}=0,\label{N=3b}\\
& \mu v_{3,t}+3\rho v_{3,x}+v_{3,xxx}=0,\label{N=3c}
\end{align}
where
\begin{equation}
\rho= 2(c_{12}v_1v_2+c_{13}v_1v_3+c_{23}v_2v_3).
\end{equation}
The corresponding Hirota bilinear form is
\begin{align}
&(\mu D_t+D_x^3)\{g_i\cdot f\}=0,\quad i=1, 2, 3, \\
&D_x^2\{f\cdot f\}= 2(c_{12}g_1g_2+c_{13}g_1g_3+c_{23}g_2g_3).
\end{align}

\subsubsection{One-Soliton Solution of N=3 Coupled HI-mKdV System}

Similar to $N=2$ case we take the solution given by (\ref{g_if}) with the pfaffian elements (\ref{pf1})-(\ref{pf3}) under the condition (\ref{pf4}).
The solution which has one-soliton for every $\displaystyle v_i=\frac{g_i}{f}$, $i=1, 2, 3$, i.e., $M_j=1$, $j=1, 2, 3$ so $L=M_1+M_2+M_3=3$ with
$B_j=\{b_j\}$, $j=1, 2, 3$  is expressed by
\begin{align}
&g_i=\mathrm{pf}(d_0,a_1,a_2,a_3,b_1,b_2,b_3,\beta_i),\, i=1,2,3,\\
&f=\mathrm{pf}(a_1,a_2,a_3,b_1,b_2,b_3),
\end{align}
which are explicitly given as
\begin{align}
& g_1=-e^{\theta_1}-\alpha_{13}\alpha_{12}\alpha_{23}\beta_{23}e^{\theta_1+\theta_2+\theta_3},\\
& g_2=-e^{\theta_2}+\alpha_{13}\alpha_{12}\alpha_{23}\beta_{13}e^{\theta_1+\theta_2+\theta_3},\\
& g_3=-e^{\theta_3}-\alpha_{13}\alpha_{12}\alpha_{23}\beta_{12}e^{\theta_1+\theta_2+\theta_3},\\
&f=-1-\alpha_{12}\beta_{12}e^{\theta_1+\theta_2}-\alpha_{13}\beta_{13}e^{\theta_1+\theta_3}-\alpha_{23}\beta_{23}e^{\theta_2+\theta_3},
\end{align}
where $\theta_i=k_ix-\frac{k_i^3}{\mu}t+\delta_i$, $\displaystyle \alpha_{ij}=\frac{k_i-k_j}{k_i+k_j}$, and $\displaystyle \beta_{ij}=\frac{c_{ij}}{k_i^2-k_j^2}$ for
$i,j=1,2,3$. Here $k_i, \delta_i$, $i=1, 2, 3$ are arbitrary constants.

\subsection{Local and Nonlocal Reductions for N=3}

The coupled HI-mKdV system for $N=3$ given by (\ref{N=3a})-(\ref{N=3c}) has eight consistent reductions; four of them are local and the others are nonlocal.
To obtain one-soliton solution of the reduced equations one can use Type 1 and Type 2 approaches given in \cite{GurPek2}. Here we will only use Type 1.

\subsubsection{Local Reductions for N=3}

We have four different local reductions for (\ref{N=3a})-(\ref{N=3c}).

\noindent \textbf{i.} $v_3=a_3+a_2v_2+a_1v_1$, $a_i$, $i=1, 2, 3$ are constants. When we use this reduction,
the system (\ref{N=3a})-(\ref{N=3c}) consistently reduces to
\begin{align}
&\mu v_{1,t}+6((c_{12}+c_{13}a_2+c_{23}a_1)v_1v_2+c_{13}a_1v_1^2+c_{23}a_2v_2^2+c_{13}a_3v_1+c_{23}a_3v_2)v_{1,x}+v_{1,xxx}=0,\nonumber\\
&\mu v_{2,t}+6((c_{12}+c_{13}a_2+c_{23}a_1) v_1v_2+c_{13}a_1v_1^2+c_{23}a_2v_2^2+c_{13}a_3v_1+c_{23}a_3v_2)v_{2,x}+v_{2,xxx}=0, \label{localiN=3}
\end{align}
without any additional condition.  If we follow the Type 1 approach, the constraints that the parameters of the one-soliton solution of the system (\ref{localiN=3}) are obtained from
\begin{equation}
\frac{g_3}{f}=a_3+a_2\frac{g_2}{f}+a_1\frac{g_1}{f} \Rightarrow g_3=a_3f+a_2g_2+a_1g_1
\end{equation}
 as
\begin{equation}
1)\, a_3=0,\quad 2)\, k_1=k_2=k_3,\quad 3)\, e^{\delta_3}=a_2e^{\delta_2}+a_1e^{\delta_1}.
\end{equation}
In this case we have $v_2=\alpha v_1$, where $\alpha=e^{\delta_2-\delta_1}$, and the system (\ref{localiN=3}) becomes a single equation, usual mKdV equation,
\begin{equation}\label{reducedeqlocaliN=3}
\mu v_{1,t}+6\zeta v_1^2 v_{1,x}+v_{1,xxx}=0,
\end{equation}
where $\zeta=(c_{12}+c_{13}a_2+c_{23}a_1)\alpha+c_{13}a_1+c_{23}a_2\alpha^2$. We obtain the one-soliton solution of
(\ref{reducedeqlocaliN=3}) as
\begin{equation}\displaystyle
v_1=\frac{e^{k_1x-\frac{k_1^3}{\mu}t+\delta_1}}{1+\frac{1}{4k_1^2}[c_{13}a_1e^{2\delta_1}+c_{23}a_2e^{2\delta_2}
+(c_{12}+c_{13}a_2+c_{23}a_1)e^{\delta_1+\delta_2}]e^{2k_1x-\frac{2k_1^3}{\mu}t}}.
\end{equation}

\vspace{0.7cm}

\noindent \textbf{ii.} $v_3=a_3+a_2\bar{v}_2+a_1\bar{v}_1$, $a_i$, $i=1, 2, 3$ are constants. If we apply this reduction to
the system (\ref{N=3a})-(\ref{N=3c}), it consistently reduces if
\begin{equation}
\mu=\bar{\mu},\quad  \rho=\bar{\rho}.
\end{equation}
The relation $\rho=\bar{\rho}$ is satisfied if
\begin{equation}\label{condlocaliN=3}
1)\, c_{12}=0,\, 2)\, a_3=0,\, 3)\, c_{13}a_2=\bar{c}_{23}\bar{a}_1,\, 4)\, c_{13}a_1=\bar{c}_{13}\bar{a}_1,\,
5)\,c_{23}a_2=\bar{c}_{23}\bar{a}_2.
\end{equation}
The reduced system is
\begin{align}
&\mu v_{1,t}+6(c_{13}a_2 v_1\bar{v}_2+c_{23}a_1\bar{v}_1v_2+c_{13}a_1|v_1|^2+c_{23}a_2|v_2|^2)v_{1,x}+v_{1,xxx}=0,\nonumber\\
&\mu v_{2,t}+6(c_{13}a_2 v_1\bar{v}_2+c_{23}a_1\bar{v}_1v_2+c_{13}a_1|v_1|^2+c_{23}a_2|v_2|^2)v_{2,x}+v_{2,xxx}=0,\label{localiiN=3}
\end{align}
with the conditions (\ref{condlocaliN=3}) satisfied and $\mu \in \mathbb{R}$. If we use the Type 1 approach to find one-soliton solution of the above system, we obtain the constraints that the parameters of the one-soliton solution of the system (\ref{localiiN=3}) from
\begin{equation}
\frac{g_3}{f}=a_2\frac{\bar{g}_2}{\bar{f}}+a_1\frac{\bar{g}_1}{\bar{f}} \Rightarrow f=\bar{f}, \, g_3=a_2\bar{g}_2+a_1\bar{g}_1,
\end{equation}
as
\begin{equation}
1)\, k_3=\bar{k}_2=\bar{k}_1,\quad 2)\, e^{\delta_3}=a_2e^{\bar{\delta}_2}+a_1e^{\bar{\delta}_1}.
\end{equation}
In this case we have $v_2=e^{\delta_2-\delta_1} v_1$ and the system (\ref{localiiN=3}) becomes a single cmKdV equation,
\begin{equation}\label{reducedeqlocaliiN=3}
\mu v_{1,t}+6[c_{13}(\bar{\alpha}a_2+a_1)+c_{23}(\alpha a_1+\alpha^2 a_2)]|v_1|^2 v_{1,x}+v_{1,xxx}=0.
\end{equation}
One-soliton solution of the equation
(\ref{reducedeqlocaliiN=3}) is
\begin{equation}\displaystyle
v_1=\frac{e^{k_1x-\frac{k_1^3}{\mu}t+\delta_1}}{1+\frac{1}{(k_1+\bar{k}_1)^2}(c_{13}e^{\delta_1}+c_{23}e^{\delta_2})(a_2e^{\bar{\delta}_2}
+a_1e^{\bar{\delta}_1})e^{(k_1+\bar{k}_1)x-\frac{(k_1^3+\bar{k}_1^3)}{\mu}t}}.
\end{equation}

\vspace{0.7cm}

\noindent \textbf{iii.} $v_3=a_1+a_2v_1$, $v_2=b_1+b_2v_1$ $a_i, b_i$, $i=1, 2, 3$ are constants. When we apply this reduction to
the system (\ref{N=3a})-(\ref{N=3c}) it consistently reduces to the equation
\begin{equation}\label{localiiiN=3}
\mu v_{1,t}+6([c_{12}b_2+c_{13}a_2+c_{23}a_2b_2]v_1^2+[c_{12}b_1+c_{13}a_1+c_{23}a_1b_2+c_{23}a_2b_1]v_1+c_{23}a_1b_1)v_{1,x}+v_{1,xxx}=0,
\end{equation}
without any additional condition. When we use the Type 1 approach, we obtain the constraints that the parameters of the one-soliton solution of the equation (\ref{localiiiN=3}) are obtained from
\begin{equation}
\frac{g_3}{f}=a_1+a_2\frac{g_1}{f},\, \frac{g_2}{f}=b_1+b_2\frac{g_1}{f}  \Rightarrow g_3=a_1f+a_2g_1,\, g_2=b_1f+b_2g_1
\end{equation}
 as
\begin{equation}
1)\, a_1=b_1=0,\, 2)\, k_1=k_2=k_3,\, 3)\, e^{\delta_3}=a_2e^{\delta_1},\, 4)\, e^{\delta_2}=b_2e^{\delta_1},
\, 5)\, c_{12}=-a_2c_{23}, \, 6)\, c_{13}=-b_2c_{23}.
\end{equation}
In this case the equation (\ref{localiiiN=3}) becomes the usual mKdV equation,
\begin{equation}\label{reducedeqlocaliiiN=3}
\mu v_{1,t}+6c_{12}b_2 v_1^2 v_{1,x}+v_{1,xxx}=0,
\end{equation}
and its one-soliton solution is
\begin{equation}\displaystyle
v_1=\frac{e^{k_1x-\frac{k_1^3}{\mu}t+\delta_1}}{1+\frac{e^{2\delta_1}}{4k_1^2}c_{12}b_2e^{2k_1x-\frac{2k_1^3}{\mu}t}}.
\end{equation}

\vspace{0.7cm}

\noindent \textbf{iv.} $v_3=a_1+a_2v_1+a_3\bar{v}_1$, $v_2=b_1+b_2v_1+b_3\bar{v}_1$ $a_i, b_i$, $i=1, 2, 3$ are constants. If we apply this reduction to
the system (\ref{N=3a})-(\ref{N=3c}), it consistently reduces when
\begin{equation}
 \mu=\bar{\mu},\quad  \rho=\bar{\rho}.
\end{equation}
The relation $\rho=\bar{\rho}$ is satisfied if
\begin{align}
&1)\, c_{23}a_1b_1=\bar{c}_{23}\bar{a}_1\bar{b}_1,\nonumber\\
&2)\, c_{23}a_1b_3+c_{23}a_3b_1=\bar{c}_{12}\bar{b}_1+\bar{c}_{13}\bar{a}_1+\bar{c}_{23}\bar{a}_1\bar{b}_2+\bar{c}_{23}\bar{a}_2\bar{b}_1,\nonumber\\
&3)\, c_{12}b_2+c_{13}a_2+c_{23}a_2b_2=\bar{c}_{23}\bar{a}_3\bar{b}_3,\nonumber\\
&4)\, c_{12}b_3+c_{13}a_3+c_{23}a_2b_3+c_{23}a_3b_2=\bar{c}_{12}\bar{b}_3+\bar{c}_{13}\bar{a}_3+\bar{c}_{23}\bar{a}_2\bar{b}_3+\bar{c}_{23}\bar{a}_3\bar{b}_2.\label{condlocalivN=3}
\end{align}
The reduced equation is
\begin{equation}\label{localivN=3}
\mu v_{1,t}+6(\gamma_1v_1^2+\gamma_2v_1+\gamma_3\bar{v}_1^2+\gamma_4\bar{v}_1+\gamma_5v_1\bar{v}_1+\gamma_6)v_{1,x}+v_{1,xxx}=0,
\end{equation}
where
\begin{align*}
&\gamma_1=c_{12}b_2+c_{13}a_2+c_{23}a_2b_2,\,  \gamma_2=c_{12}b_1+c_{13}a_1+c_{23}a_1b_2+c_{23}a_2b_1,\, \gamma_3=c_{23}a_3b_3 \\
&\gamma_4=c_{23}a_1b_3+c_{23}a_3b_1, \gamma_5=c_{12}b_3+c_{13}a_3+c_{23}a_3b_2+c_{23}a_2b_3,\, \gamma_6=c_{23}a_1b_1,
\end{align*}
with the conditions (\ref{condlocalivN=3}) satisfied and $\mu \in \mathbb{R}$. When we use Type 1, we obtain the conditions on the
parameters of one-soliton solution of the equation (\ref{localivN=3}) as
\begin{align}
&1)\, a_1=b_1=0,\,\,\, 2)\, k_1=k_2=k_3, k_i\in \mathbb{R}, i=1, 2, 3, \,\,\, 3)\, e^{\delta_2}=b_2e^{\delta_1}+b_3e^{\bar{\delta}_1},\nonumber \\
&4)\, e^{\delta_3}=a_2e^{\delta_1}+a_3e^{\bar{\delta}_1},\,\,\, 5)\, c_{ij}e^{\delta_i+\delta_j}=\bar{c}_{ij}e^{\bar{\delta}_i+\bar{\delta}_j},\, i,j= 1, 2, 3.
\end{align}
With the above conditions, the equation (\ref{localivN=3}) becomes a cmKdV equation
\begin{equation}\label{reducedlocalivN=3}
\mu v_{1,t}+6(\gamma_1v_1^2+\gamma_3\bar{v}_1^2+\gamma_5|v_1|^2)v_{1,x}+v_{1,xxx}=0,
\end{equation}
and one-soliton solution of (\ref{reducedlocalivN=3}) is
\begin{equation}
v_1=e^{k_1x-\frac{k_1^3}{\mu}t+\delta_1}.
\end{equation}

\subsubsection{Nonlocal Reductions for N=3}

We have four different nonlocal reductions for (\ref{N=3a})-(\ref{N=3c}).

\noindent \textbf{i.} $v_3(t,x)=a_3+a_2v_2(\varepsilon_1t,\varepsilon_2 x)+a_1v_1(\varepsilon_1t,\varepsilon_2 x)=a_3+a_2v_2^{\varepsilon}+a_1v_1^{\varepsilon}$, $\varepsilon_j^2=1$, $j=1, 2$, and $a_i$, $i=1, 2, 3$ are constants. When we use this reduction
in (\ref{N=3b}) for consistency of reduction we get the following conditions:
\begin{equation}
1)\, \varepsilon_1\varepsilon_2=1,\quad 2)\, a_3=0,\quad 3)\, c_{12}=0,\quad 4)\, c_{13}a_2=c_{23}a_1.
\end{equation}
Therefore to have a nonlocal equation, there is only one possibility; $(\varepsilon_1,\varepsilon_2)=(-1,-1)$. The reduced equation is ST-symmetric nonlocal mKdV system,
\begin{align}
&\mu v_{1,t}(t,x)+6[c_{13}v_1(t,x)+c_{23}v_2(t,x)][a_1v_1(-t,-x)+a_2v_2(-t,-x)]v_{1,x}(t,x)+v_{1,xxx}(t,x)=0,\nonumber\\
&\mu v_{2,t}(t,x)+6[c_{13}v_1(t,x)+c_{23}v_2(t,x)][a_1v_1(-t,-x)+a_2v_2(-t,-x)]v_{2,x}(t,x)+v_{2,xxx}(t,x)=0.\label{nonlocaliN=3}
\end{align}
If we use the Type 1 approach, the constraints that the parameters of the one-soliton solution of the system (\ref{nonlocaliN=3}) are obtained from
\begin{equation}
\frac{g_3}{f}=a_2\frac{g_2^{\epsilon}}{f^{\epsilon}}+a_1\frac{g_1^{\epsilon}}{f^{\epsilon}} \Rightarrow f=f^{\epsilon},\, g_3=a_2g_2^{\epsilon}+a_1g_1^{\epsilon}
\end{equation}
 as
\begin{equation}
1)\, k_1=k_2=-k_3,\quad 2)\, e^{\delta_3}=a_2e^{\delta_2}+a_1e^{\delta_1},\quad 3)\, c_{12}=c_{13}=c_{23}=0.
\end{equation}
The above conditions yields that $v_2=e^{\delta_2-\delta_1}v_1$. The nonlocal system (\ref{nonlocaliN=3}) reduces to local linear
equation
\begin{equation}
\mu v_{1,t}(t,x)+v_{1,xxx}(t,x)=0,
\end{equation}
and one-soliton solution of the above equation is
\begin{equation}
v_1=e^{k_1x-\frac{k_1^3}{\mu}t+\delta_1}.
\end{equation}

\vspace{0.7cm}

\noindent \textbf{ii.} $v_3(t,x)=a_3+a_2\bar{v}_2(\varepsilon_1t,\varepsilon_2 x)+a_1\bar{v}_1(\varepsilon_1t,\varepsilon_2 x)=a_3+a_2\bar{v}_2^{\varepsilon}+a_1\bar{v}_1^{\varepsilon}$, $\varepsilon_j^2=1$, $j=1, 2$, and $a_i$, $i=1, 2, 3$ are constants. Applying this reduction to (\ref{N=3c}) gives the constraints on the parameters as
\begin{equation}\label{condnonlocaliiN=3}
1)\, \mu=\bar{\mu}\varepsilon_1\varepsilon_2,\, 2)\, a_3=0,\, 3)\, c_{12}=0,\, 4)\, c_{13}a_2=\bar{c}_{23}\bar{a}_1,\, 5)\, c_{13}a_1=\bar{c}_{13}\bar{a}_1,\,
6)\, c_{23}a_2=\bar{c}_{23}\bar{a}_2
\end{equation}
for consistent reduction. Hence we have
\begin{align}
&\mu v_{1,t}(t,x)+6[c_{13}v_1(t,x)+c_{23}v_2(t,x)][a_1\bar{v}_1(\epsilon_1t,\epsilon_2x)+a_2\bar{v}_2(\epsilon_1t,\epsilon_2x)]v_{1,x}(t,x)
+v_{1,xxx}(t,x)=0,\nonumber\\
&\mu v_{2,t}(t,x)+6[c_{13}v_1(t,x)+c_{23}v_2(t,x)][a_1\bar{v}_1(\epsilon_1t,\epsilon_2x)+a_2\bar{v}_2(\epsilon_1t,\epsilon_2x)]v_{2,x}(t,x)
+v_{2,xxx}(t,x)=0.\label{NONLOCALiiN=3}
\end{align}
According to $(\epsilon_1,\epsilon_2)=\{(-1,1),(1,-1),(-1,-1)\}$ we have
three different nonlocal cmKdV systems:\\

\noindent \textbf{a.} \textbf{T-Symmetric Nonlocal CMKdV System:}
\begin{align}
&\mu v_{1,t}(t,x)+6[c_{13}v_1(t,x)+c_{23}v_2(t,x)][a_1\bar{v}_1(-t,x)+a_2\bar{v}_2(-t,x)]v_{1,x}(t,x)+v_{1,xxx}(t,x)=0,\nonumber\\
&\mu v_{2,t}(t,x)+6[c_{13}v_1(t,x)+c_{23}v_2(t,x)][a_1\bar{v}_1(-t,x)+a_2\bar{v}_2(-t,x)]v_{2,x}(t,x)+v_{2,xxx}(t,x)=0,\label{TsymmN=3}
\end{align}
where $\mu=-\bar{\mu}$ and the conditions (\ref{condnonlocaliiN=3}) hold.\\

\noindent \textbf{b.} \textbf{S-Symmetric Nonlocal CMKdV System:}
\begin{align}
&\mu v_{1,t}(t,x)+6[c_{13}v_1(t,x)+c_{23}v_2(t,x)][a_1\bar{v}_1(t,-x)+a_2\bar{v}_2(t,-x)]v_{1,x}(t,x)+v_{1,xxx}(t,x)=0,\nonumber\\
&\mu v_{2,t}(t,x)+6[c_{13}v_1(t,x)+c_{23}v_2(t,x)][a_1\bar{v}_1(t,-x)+a_2\bar{v}_2(t,-x)]v_{2,x}(t,x)+v_{2,xxx}(t,x)=0,\label{SsymmN=3}
\end{align}
where $\mu=-\bar{\mu}$ and the conditions (\ref{condnonlocaliiN=3}) hold.\\

\noindent \textbf{c.} \textbf{ST-Symmetric Nonlocal CMKdV System:}
\begin{align}
&\mu v_{1,t}(t,x)+6[c_{13}v_1(t,x)+c_{23}v_2(t,x)][a_1\bar{v}_1(-t,-x)+a_2\bar{v}_2(-t,-x)]v_{1,x}(t,x)+v_{1,xxx}(t,x)=0,\nonumber\\
&\mu v_{2,t}(t,x)+6[c_{13}v_1(t,x)+c_{23}v_2(t,x)][a_1\bar{v}_1(-t,-x)+a_2\bar{v}_2(-t,-x)]v_{2,x}(t,x)+v_{2,xxx}(t,x)=0,\label{STsymmN=3}
\end{align}
where $\mu=\bar{\mu}$ and the conditions (\ref{condnonlocaliiN=3}) hold.\\

When we apply the Type 1 approach, the constraints that the parameters of the one-soliton solution of the system (\ref{NONLOCALiiN=3}) are obtained
 as
\begin{equation}
1)\, k_3=\epsilon_2\bar{k}_2=\epsilon_2\bar{k}_1,\quad 2)\, e^{\delta_3}=a_2e^{\bar{\delta}_2}+a_1e^{\bar{\delta}_1},
\quad 3)\, c_{13}e^{\delta_1+\delta_3}=\bar{c_{13}}e^{\bar{\delta}_1+\bar{\delta}_3}, \quad 4)\, c_{23}e^{\delta_2+\delta_3}=\bar{c_{23}}e^{\bar{\delta}_2+\bar{\delta}_3}.
\end{equation}
In this case $v_2(t,x)=e^{\delta_2-\delta_1}v_1(t,x)$ and the system (\ref{NONLOCALiiN=3}) reduces to a single nonlocal cmKdV equation \cite{GurPek2}
\begin{equation}
\mu v_1(t,x)+6(c_{13}+c_{23}e^{\delta_2-\delta_1})(a_1+a_2e^{\bar{\delta}_2-\bar{\delta}_1})v_1(t,x)
\bar{v}_1(\epsilon_1t,\epsilon_2x)v_{1,x}(t,x)+v_{1,xxx}(t,x)=0.
\end{equation}
One-soliton solution of the above equation is
\begin{equation}
v_1=\frac{e^{k_1x-\frac{k_1^3}{\mu}t+\delta_1}}{1+\frac{1}{(k_1+\epsilon_2\bar{k}_1)^2}[c_{13}e^{\delta_1}+c_{23}e^{\delta_2}]
[a_2e^{\bar{\delta}_2}+a_1e^{\bar{\delta}_1}]e^{(k_1+\epsilon_2\bar{k}_1)x-\frac{(k_1^3+\epsilon_2\bar{k}_1^3)}{\mu}t}},
\end{equation}
with the conditions (\ref{condnonlocaliiN=3}) hold.

\vspace{0.7cm}

\noindent \textbf{iii.} $v_3(t,x)=a_1+a_2v_1(t,x)+a_3v_1(\varepsilon_1t,\varepsilon_2 x)=a_1+a_2v_1+a_3v_1^{\varepsilon}$, $v_2(t,x)=b_1+b_2v_1(t,x)+b_3v_1(\varepsilon_1t,\varepsilon_2 x)=b_1+b_2v_1+b_3v_1^{\varepsilon}$ where $a_i, b_i$, $i=1, 2, 3$ are constants, and $\varepsilon_j^2=1$, $j=1, 2$. Applying this reduction to (\ref{N=3b}) and (\ref{N=3c}) gives the constraints on the parameters as
\begin{align}\label{condnonlocaliiiN=3}
&1)\, \varepsilon_1\varepsilon_2=1,\nonumber\\
&2)\, c_{12}b_1+c_{13}a_1+c_{23}a_2b_1+c_{23}a_1b_2=c_{23}a_1b_3+c_{23}a_3b_1,\nonumber\\
&3)\, c_{12}b_2+c_{13}a_2+c_{23}a_2b_2=c_{23}a_3b_3,
\end{align}
for consistent reduction. Here we have one possibility; $(\varepsilon_1,\varepsilon_2)=(-1,-1)$. The reduced equation is
\begin{align}\label{nonlocaliiiN=3}
&\mu v_{1,t}(t,x)+6[(c_{12}b_2+c_{13}a_2+c_{23}a_2b_2)v_1^2(t,x)+(c_{12}b_1+c_{13}a_1+c_{23}a_2b_1+c_{23}a_1b_2)v_1(t,x)\nonumber\\
&+c_{23}a_3b_3v_1^2(-t,-x)+(c_{23}a_1b_3+c_{23}a_3b_1)v_1(-t,-x)\nonumber\\
&+(c_{12}b_3+c_{13}a_3+c_{23}a_2b_3+c_{23}a_3b_2)v_1(t,x)v_1(-t,-x)+c_{23}a_1b_1]v_{1,x}(t,x)+v_{1,xxx}(t,x)=0,
\end{align}
with the conditions (\ref{condnonlocaliiiN=3}) hold. If we use the Type 1 approach to obtain the one-soliton solution of the equation we get the
following constraints:
\begin{equation}
1)\, a_1=a_3=b_1=b_2=0,\, 2)\, k_1=-k_2=k_3,\, 3)\, c_{12}=c_{13}=c_{23}=0,\, 4)\, e^{\delta_3}=a_2e^{\delta_1},\, 5)\, e^{\delta_2}=b_3e^{\delta_1}.
\end{equation}
Hence the equation (\ref{nonlocaliiiN=3}) reduces to the local linear equation
\begin{equation}
\mu v_{1,t}(t,x)+v_{1,xxx}(t,x)=0,
\end{equation}
and one-soliton solution of this equation is
\begin{equation}
v_1=e^{k_1x-\frac{k_1^3}{\mu}t+\delta_1}.
\end{equation}

\vspace{0.7cm}

\noindent \textbf{iv.} $v_3(t,x)=a_1+a_2v_1(t,x)+a_3\bar{v}_1(\varepsilon_1t,\varepsilon_2 x)=a_1+a_2v_1+a_3\bar{v}_1^{\varepsilon}$, $v_2(t,x)=b_1+b_2v_1(t,x)+b_3\bar{v}_1(\varepsilon_1t,\varepsilon_2 x)=b_1+b_2v_1+b_3\bar{v}_1^{\varepsilon}$ where $a_i, b_i$, $i=1, 2, 3$ are constants, and $\varepsilon_j^2=1$, $j=1, 2$. When we apply this reduction to (\ref{N=3b}) and (\ref{N=3c}) we obtain the constraints on the parameters as
\begin{align}\label{condnonlocalivN=3}
&1)\, \bar{\mu}\varepsilon_1\varepsilon_2=\mu,\nonumber\\
&2)\, c_{12}b_1+c_{13}a_1+c_{23}a_2b_1+c_{23}a_1b_2=\bar{c}_{23}\bar{a}_1\bar{b}_3+\bar{c}_{23}\bar{a}_3\bar{b}_1,\nonumber\\
&3)\, c_{12}b_2+c_{13}a_2+c_{23}a_2b_2=\bar{c}_{23}\bar{a}_3\bar{b}_3,\nonumber\\
&4)\, c_{12}b_3+c_{13}a_3+c_{23}a_3b_2+c_{23}a_2b_3=\bar{c}_{12}\bar{b}_3+\bar{c}_{13}\bar{a}_3+
\bar{c}_{23}\bar{a}_3\bar{b}_2+\bar{c}_{23}\bar{a}_2\bar{b}_3,\nonumber\\
&5\, c_{23}a_1b_1=\bar{c}_{23}\bar{a}_1\bar{b}_1,
\end{align}
for consistent reduction. Hence we have a single nonlocal cmKdV equation
\begin{align}\label{nonlocalivN=3}
&\mu v_{1,t}(t,x)+6[\bar{c}_{23}\bar{a}_3\bar{b}_3v_1^2(t,x)+(c_{12}b_1+c_{13}a_1+c_{23}a_2b_1+c_{23}a_1b_2)v_1(t,x)\nonumber\\
&+c_{23}a_3b_3\bar{v}_1^2(\varepsilon_1t,\varepsilon_2 x)+(c_{23}a_1b_3+c_{23}a_3b_1)\bar{v}_1(\varepsilon_1t,\varepsilon_2 x)\nonumber\\
&+(c_{12}b_3+c_{13}a_3+c_{23}a_2b_3+c_{23}a_3b_2)v_1(t,x)\bar{v}_1(\varepsilon_1t,\varepsilon_2 x)+c_{23}a_1b_1]v_{1,x}(t,x)+v_{1,xxx}(t,x)=0.
\end{align}

\noindent Therefore we have three different nonlocal cmKdV equations:\\

\noindent \textbf{a.} \textbf{T-Symmetric Nonlocal CMKdV Equation:}
\begin{align}
&\mu v_{1,t}(t,x)+6[\bar{c}_{23}\bar{a}_3\bar{b}_3v_1^2(t,x)+(c_{12}b_1+c_{13}a_1+c_{23}a_2b_1+c_{23}a_1b_2)v_1(t,x)\nonumber\\
&+c_{23}a_3b_3\bar{v}_1^2(-t,x)+(c_{23}a_1b_3+c_{23}a_3b_1)\bar{v}_1(-t,x)\nonumber\\
&+(c_{12}b_3+c_{13}a_3+c_{23}a_2b_3+c_{23}a_3b_2)v_1(t,x)\bar{v}_1(-t,x)+c_{23}a_1b_1]v_{1,x}(t,x)+v_{1,xxx}(t,x)=0,
\end{align}
where $\mu=-\bar{\mu}$ and the conditions (\ref{condnonlocalivN=3}) hold.\\

\noindent \textbf{b.} \textbf{S-Symmetric Nonlocal CMKdV  Equation:}
\begin{align}
&\mu v_{1,t}(t,x)+6[\bar{c}_{23}\bar{a}_3\bar{b}_3v_1^2(t,x)+(c_{12}b_1+c_{13}a_1+c_{23}a_2b_1+c_{23}a_1b_2)v_1(t,x)\nonumber\\
&+c_{23}a_3b_3\bar{v}_1^2(t,-x)+(c_{23}a_1b_3+c_{23}a_3b_1)\bar{v}_1(t,-x)\nonumber\\
&+(c_{12}b_3+c_{13}a_3+c_{23}a_2b_3
+c_{23}a_3b_2)v_1(t,x)\bar{v}_1(t,-x)+c_{23}a_1b_1]v_{1,x}(t,x)+v_{1,xxx}(t,x)=0,
\end{align}
where $\mu=-\bar{\mu}$ and the conditions (\ref{condnonlocalivN=3}) hold.\\

\noindent \textbf{c.} \textbf{ST-Symmetric Nonlocal CMKdV  Equation:}
\begin{align}
&\mu v_{1,t}(t,x)+6[\bar{c}_{23}\bar{a}_3\bar{b}_3v_1^2(t,x)+(c_{12}b_1+c_{13}a_1+c_{23}a_2b_1+c_{23}a_1b_2)v_1(t,x)\nonumber\\
&+c_{23}a_3b_3\bar{v}_1^2(-t,-x)+(c_{23}a_1b_3+c_{23}a_3b_1)\bar{v}_1(-t,-x)\nonumber\\
&+(c_{12}b_3+c_{13}a_3+c_{23}a_2b_3+c_{23}a_3b_2)v_1(t,x)\bar{v}_1(-t,-x)+c_{23}a_1b_1]v_{1,x}(t,x)+v_{1,xxx}(t,x)=0,
\end{align}
where $\mu=\bar{\mu}$ and the conditions (\ref{condnonlocalivN=3}) hold.\\

\noindent By using Type 1 approach, we obtain the constraints that the parameters of one-soliton solution of the equation
(\ref{nonlocalivN=3}) as
\begin{align}
&1)\, a_1=b_1=0,\quad 2)\, k_1=k_2=k_3=\varepsilon_2\bar{k}_1,\quad  3)\, e^{\delta_2}=b_2e^{\delta_1}+b_3e^{\bar{\delta}_1},\nonumber\\
&4)\, e^{\delta_3}=a_2e^{\delta_1}+a_3e^{\bar{\delta}_1},\quad  5)\, c_{ij}e^{\delta_i+\delta_j}=\bar{c}_{ij}e^{\bar{\delta}_i+\bar{\delta}_j},\, i, j=1, 2, 3.
\end{align}
With these conditions, the equation (\ref{nonlocalivN=3}) becomes the following nonlocal cmKdV equation:
\begin{align}\label{reducednonlocalivN=3}
&\mu v_{1,t}(t,x)+6[(c_{12}b_2+c_{13}a_2+c_{23}a_2b_2)v_1^2(t,x)+c_{23}a_3b_3\bar{v}_1^2(\varepsilon_1t,\varepsilon_2 x)\nonumber\\
&+(c_{12}b_3+c_{13}a_3+c_{23}a_2b_3+c_{23}a_3b_2)v_1(t,x)\bar{v}_1(\varepsilon_1t,\varepsilon_2 x)]v_{1,x}(t,x)+v_{1,xxx}(t,x)=0.
\end{align}
One-soliton solution of the equation (\ref{reducednonlocalivN=3}) is
\begin{equation}
v_1(t,x)=e^{k_1x-\frac{k_1^3}{\mu}t+\delta_1}.
\end{equation}

 \begin{rem}
For $N=3$ we have reductions to system of two equations, but if we require Type 1 one-soliton solution, they reduce to a single equation.
 \end{rem}

\section{Acknowledgment}
  The author thanks M. G\"{u}rses for helpful discussions and valuable suggestions. This work is partially supported by the Scientific
and Technological Research Council of Turkey (T\"{U}B\.{I}TAK).

\end{document}